\documentclass[twocolumn]{jpsj2}

\title{
The Kondo Lattice Model in Infinite Dimensions I. Formalism
}

\author{%
Junya \textsc{Otsuki}\thanks{E-mail address: otsuki@cmpt.phys.tohoku.ac.jp}, 
Hiroaki \textsc{Kusunose}$^1$ and Yoshio \textsc{Kuramoto}
}

\inst{%
Department of Physics, Tohoku University, Sendai 980-8578\\
$^1$Department of Physics, Ehime University, Matsuyama 790-8577
}

\recdate{\today}

\abst{%
A Green-function formalism for 
the Kondo lattice model is presented, which is designed to be combined with the dynamical mean-field theory.
With use of Wick's theorem only for conduction electrons, 
dynamical quantities are represented in terms of the $t$-matrix and its two-particle generalizations.
By taking the high-frequency limit of the $t$-matrix
with respect to a part of the fermion frequencies, one obtains
dynamical correlation functions of localized moments. 
Explicit examples of calculational steps are provided with use of the continuous-time quantum Monte Carlo method.
}

\kword{%
equation of motion, two-particle Green function, generalized $t$-matrix, 
Kondo lattice model, Coqblin-Schrieffer lattice model, Anderson lattice model, 
dynamical mean-field theory (DMFT), continuous-time quantum Monte Carlo method (CT-QMC)
}

\begin{document}
\maketitle

\section{Introduction}

In this paper, we present a formalism which is designed to be applicable to 
the Kondo lattice and related models with localized degrees of freedom.
The Kondo lattice is the simplest model to investigate rich consequence of the collective Kondo effect such as heavy electrons, and competition with magnetically ordered states.
In higher dimensions, the dynamical mean-field theory (DMFT) is the most effective and the simplest approach.\cite{Georges}
If one tries to apply the DMFT to the Kondo lattice, however, a difficulty arises because the localized moments cannot be dealt with by the ordinary Green functions. 
This difficulty results from the special situation of the strong correlation limit where the charge degree of freedom has been eliminated. 
Namely the fermionic Green function cannot be defined for the localized degrees of freedom.
In this sense, the Anderson lattice is simpler since the Green function of localized electrons is well defined.  However, the strong correlation limit of the Anderson lattice is numerically more difficult because the high-energy intermediate states are essential in addition to singly occupied states for each site. 
Therefore,  we pursue a formalism which allows one to perform highly accurate numerical calculation dealing only with low-energy localized states.

In 
addressing to the localized limit, it is instructive to 
compare with the local Fermi liquid theory, which describes low-temperature properties of the Kondo problem. 
Nozi\`eres has described the low-energy excitations 
in terms of the phase shift of conduction electrons near the Fermi level\cite{Nozieres}.
The phase shift 
contains all information of the impurity scattering caused by the exchange interaction.
By expanding the phase shift with respect to the quasi-particle distribution, 
one obtains the Wilson ratio as well as the transport properties at low temperatures. 
The same consequence can be derived by means of the perturbative approach of the Anderson model\cite{Yamada}.
Our approach is a microscopic version of the phase-shift theory 
in the sense that we make maximum use of the viewpoint from conduction electrons.

The impurity $t$-matrix corresponds to the product of the $4f$-electron Green function and the hybridization squared in the Anderson model.
Even the localized and strong-correlation limit keeps the $t$-matrix finite.
Hence, with use of the impurity $t$-matrix, a unified treatment of single-particle properties is possible for the Anderson and the Kondo-type models.
In the same spirit, we will introduce a ``generalized $t$-matrix'' to describe two-particle correlations due to the local interactions.
It will be demonstrated that the generalized $t$-matrix gives the susceptibility of the localized moments both in the Anderson and Kondo-type models.

The ideas of the impurity $t$-matrix and the generalized $t$-matrix are also applicable to the periodic model.
With the DMFT, the Kondo lattice model is mapped to the corresponding effective impurity model. 
The effective bath is determined through the impurity $t$-matrix in a self-consistent manner.
Furthermore, we will derive spatial correlations of the localized moments within the DMFT.
The procedure which uses the Bethe-Salpeter equation with the local vertex of the effective impurity\cite{Jarrell-Hubbard, Jarrell-AL} will be extended to the Kondo lattice with use of the generalized $t$-matrix.

This paper is organized as follows.
We present impurity and periodic models in the next section. 
In \S3, the $t$-matrix is extended to the two-particle Green function with use of the equations of motions.
The DMFT self-consistent equations are given for the localized models in \S\ref{sec:dmft}. 
We further derive spatial correlations of localized moments within the DMFT.
In \S5, we present how to evaluate the generalized $t$-matrix in the continuous-time quantum Monte Carlo method (CT-QMC).
An example of the generalized $t$-matrix is explicitly given in \S6.
We finally summarize in \S7.

\section{Models}
\label{sec:model}



The Anderson lattice model has been most widely used to discuss properties of $4f$-electrons systems.
We consider a $N$-fold degenerate band represented by
\begin{align}
	H_{\rm c} = \sum_{\mib{k}\alpha} \epsilon_{\mib{k}} c_{\mib{k}\alpha}^{\dag} c_{\mib{k}\alpha},
\end{align}
where $c_{\mib{k}\alpha}^{\dag}$ and $c_{\mib{k}\alpha}$ are the creation and annihilation operators for the conduction electron with $\alpha$-th component, respectively. 
The $N$-fold degenerate Anderson lattice model is given by the following Hamiltonian:
\begin{align}
	H_{\rm 
	AL} &= H_{\rm c}
	+ \epsilon_f \sum_{i\alpha} f^{\dag}_{i\alpha} f_{i\alpha}
	+ V \sum_{i\alpha} (f^{\dag}_{i\alpha} c_{i\alpha} + c^{\dag}_{i\alpha} f_{i\alpha}) \nonumber \\
	&+ U \sum_{i, \langle \alpha \alpha' \rangle}  f^{\dag}_{i\alpha} f_{i\alpha}  f^{\dag}_{i\alpha'} f_{i\alpha'},
\label{eq:H_AL}
\end{align}
where $f_{i\alpha}^{\dag}$ and $f_{i\alpha}$ creates and annihilates the localized electron at $i$-site, respectively.
The sum over $i$ is to be taken over all the $4f$ sites, and $\langle \alpha \alpha' \rangle$ denotes a pair of different components.

By eliminating the charge degree of freedom, we derive effective models in the localized limit.
We first exclude double and higher occupations by taking $U=\infty$. 
Then taking the limit, $V \rightarrow \infty$ and $\epsilon_f \rightarrow -\infty$ with $J=-V^2/\epsilon_f$ fixed, 
the second order perturbation theory leads to the Coqblin-Schrieffer (CS) lattice model\cite{CS}
\begin{align}
	H_{\rm CSL} = H_{\rm c}
	+ J \sum_{i \alpha \alpha'} f^{\dag}_{i\alpha} f_{i\alpha'} c^{\dag}_{i\alpha'} c_{i\alpha},
\label{eq:H_CSL}
\end{align}
which describes interaction between localized and conduction electrons due to the virtual excitations to the $4f^0$ state. 
Since the intermediate $4f^0$ state is isotropic, the interaction has SU($N$) symmetry. 
If doubly occupied states are taken into account as the intermediate states, the interaction becomes anisotropic and more complicated. 

In the case of $N=2$, we can obtain a model with higher symmetry by including virtual excitations to doubly occupied state, since $4f^2$ state is isotropic as in $4f^0$. 
In the symmetric condition $U=-2\epsilon_f$, 
we take the localized limit, i.e., $V \rightarrow \infty$ and $\epsilon_f \rightarrow -\infty$ keeping the ratio $J=-2V^2/\epsilon_f$. 
Then, we obtain the Kondo lattice model
\begin{align}
	H_{\rm KL} = H_{\rm c} + J \sum_i \mib{S}_i \cdot \mib{\sigma}^{{\rm c}}_i,
\label{eq:H_KL}
\end{align}
where $\mib{S}_i$ is the localized spin at $i$-site and $\mib{\sigma}_i^{\rm c}$ is defined by $\mib{\sigma}_i^{\rm c}=\sum_{\sigma \sigma'} c_{i\sigma}^{\dag} \mib{\sigma}_{\sigma \sigma'} c_{i\sigma}$.
The interaction term is invariant under the particle-hole transformation. 
The exchange interaction relates to the CS interaction as
\begin{align}
	\mib{S}_i \cdot \mib{\sigma}_i^{\rm c}
	= \sum_{\sigma \sigma'} f^{\dag}_{i\sigma} f_{i\sigma'} c^{\dag}_{i\sigma'} c_{i\sigma}
	- \frac{1}{2} \sum_{\sigma} c^{\dag}_{i\sigma} c_{i\sigma},
\end{align}
where we have used the condition $\sum_{\sigma} f^{\dag}_{\sigma} f_{\sigma}$=1. 
The second term is the potential scattering independent of the spin component. 
The same potential for all sites are equivalent to a shift of the chemical potential, so that eq.~(\ref{eq:H_KL}) is rewritten as
\begin{align}
	H_{\rm KL} = \sum_{\mib{k}\sigma} (\epsilon_{\mib{k}} + v) c^{\dag}_{\mib{k}\sigma} c_{\mib{k}\sigma}
	+ J \sum_{i \sigma \sigma'} f^{\dag}_{i\sigma} f_{i\sigma'} c^{\dag}_{i\sigma'} c_{i\sigma},
\label{eq:H_KL2}
\end{align}
where $v=-J/2$.
The Kondo lattice model can therefore be dealt with by replacing the chemical potential $\mu$ with $\mu'=\mu-v$ in the CS lattice model. 
Hence there is essentially no difference in the physical properties between the Kondo and the CS interactions with $N=2$. 
In this paper, we employ the CS interactions, which includes $N \geq 2$.
We note that only the $t$-matrix differs in the two models since they depend on the choice of the bare Green function.

The above Hamiltonians of the lattice give the corresponding impurity models, if the index $i$ is restricted to a single site.
Hereafter, the terms `Anderson model' and `CS model' refer to both the corresponding impurity and lattice models.
We use `impurity' or `lattice' explicitly to specify either systems.

\section{Dynamics of Conduction Electrons and Generalized $t$-matrix}


In this section, we present a formalism based on the Green function to handle the Anderson and CS models in a unified way. 
In the atomic limit, it is necessary for Green-function formalism to represent all quantities in terms of the conduction electrons.
With use of equations of motions for conduction electrons, we derive dynamics of localized moments.

In the finite-temperature formalism, it is convenient to work in the grand canonical ensemble. 
To this end, we replace the kinetic energy term $H_{\rm c}$ by
\begin{align}
	H_{\rm c} - \mu N_{\rm c} = \sum_{\mib{k}\alpha} \xi_{\mib{k}} c_{\mib{k}\alpha}^{\dag} c_{\mib{k}\alpha},
\end{align}
where $\xi_{\mib{k}} = \epsilon_{\mib{k}}-\mu$ is the kinetic energy with respect to the chemical potential.

\subsection{Single-particle Green function}
We begin with the single-particle Green function prior to two-particle counterparts.
With use of the Heisenberg operator $c_{\mib{k}\alpha} (\tau) = {\rm e}^{\tau H} c_{\mib{k}\alpha} {\rm e}^{-\tau H}$, the single-particle Green function is defined by\cite{AGD, Fetter-Walecka}
\begin{align}
	G^{\rm c}_{\mib{k}\alpha \mib{k}'\alpha'}(\tau, \tau') = -\langle T_{\tau} c_{\mib{k}\alpha}(\tau) c_{\mib{k}'\alpha'}^{\dag}(\tau') \rangle,
\end{align}
where $T_{\tau}$ is the time-ordering operator, and the bracket denotes the thermal average, $\langle \cdots \rangle = \text{Tr} \{ \text{e}^{\beta (\Omega-H)} \cdots \}$ with $\text{e}^{-\beta \Omega} = \text{Tr} \text{e}^{-\beta H}$. 
In impurity systems, the site-diagonal element is given by
\begin{align}
	G^{\rm c}_{\alpha \alpha'}(\tau, \tau')=N_0^{-1} \sum_{\mib{k}\mib{k}'} G^{\rm c}_{\mib{k}\alpha \mib{k}'\alpha'}(\tau, \tau'),
\end{align}
where $N_0$ is the number of sites.
The Fourier transform is defined 
by
\begin{align}
	G^{\rm c}_{\mib{k}\alpha \mib{k}'\alpha'} ({\rm i}\epsilon_n)
	= \frac{1}{\beta} \int_0^{\beta} {\rm d}\tau \int_0^{\beta} {\rm d}\tau' 
	 G^{\rm c}_{\mib{k}\alpha \mib{k}'\alpha'}(\tau, \tau') {\rm e}^{{\rm i}\epsilon_n (\tau-\tau')}, 
\label{eq:Fourier_G}
\end{align}
where $\epsilon_n = (2n+1) \pi T$ is the Matsubara frequency for fermions. 
We express the Green function of free conduction electrons by 
$g_{\mib{k}}({\rm i}\epsilon_n) = ({\rm i}\epsilon_n - \xi_{\mib{k}})^{-1}$.

To see influences of the localized moment on the conduction electrons, we derive equations of motion for $G^{\rm c}_{\mib{k}\alpha \mib{k}'\alpha'}(\tau, \tau')$. 
To this end, we first show equations of motion for $c_{\mib{k}}(\tau)$ and $c_{\mib{k}}^{\dag}(\tau)$. 
The Heisenberg equation leads to
\begin{align}
	\frac{\partial c_{\mib{k}\alpha}(\tau)}{\partial \tau} 
	&= -\xi_{\mib{k}} c_{\mib{k}\alpha}(\tau) - j_{\mib{k}\alpha}(\tau), \nonumber \\
	\frac{\partial c_{\mib{k}\alpha}^{\dag}(\tau)}{\partial \tau} 
	&= \xi_{\mib{k}} c_{\mib{k}\alpha}^{\dag}(\tau) + j_{\mib{k}\alpha}^{\dag}(\tau),
\label{eq:motion_c}
\end{align}
where $j_{\mib{k}\alpha}$ is 
defined by $j_{\mib{k}\alpha} = [c_{\mib{k}\alpha}, H_{\rm loc}]$ with $H_{\rm loc}=H-H_{\rm c}$ being the local interaction part of the Hamiltonian\cite{Brenig-Zittartz}. 
Explicit forms of $j_{\mib{k}\alpha}$ for each model introduced in \S\ref{sec:model} are given as follows:
\begin{align}
	j_{\mib{k} \alpha} = \left\{ \begin{array}{l}
	\displaystyle{ N_0^{-1/2} V f_{\alpha}}, \qquad \text{(impurity Anderson model)} \\[0.2cm]
	\displaystyle{ Vf_{\mib{k} \alpha}},
	 \qquad \text{(Anderson lattice model, eq.~(\ref{eq:H_AL}))} \\[0.2cm]
	\displaystyle{ N_0^{-1/2} J \sum_{\alpha'}f_{\alpha'}^{\dag}f_{\alpha} c_{\alpha'}},\quad \text{(impurity CS model)} \\
	\displaystyle{ \frac{J}{N_0} \sum_{\mib{k}' \mib{q}\alpha'}f_{\mib{k}'\alpha'}^{\dag}f_{\mib{k}'+\mib{q}\alpha} c_{\mib{k}-\mib{q}\alpha'}}.\\
	 \qquad \qquad \qquad \text{(CS lattice model, eq.~(\ref{eq:H_CSL}))}
	\end{array} \right.
\label{eq:def_j}
\end{align}
By using $j_{\mib{k}\alpha}$, all models can be treated concurrently. 
Since the Kondo exchange interaction can be expressed in terms of the CS interaction as described in \S\ref{sec:model}, we employ the CS model for definiteness.
For convenient description of the impurity models, we introduce the impurity-site element $j_{\alpha}=N_0^{-1/2} \sum_{\mib{k}} j_{\mib{k}\alpha}$.
Since the interaction is local, $j_{\mib{k} \alpha}$ is independent of $\mib{k}$ in the impurity models and $j_{\alpha}$ can be simply written as $j_{\alpha}= \sqrt{N_0} j_{\mib{k}\alpha}$.

Generating an equation of motion for $G^{\rm c}_{\mib{k}\alpha \mib{k}'\alpha'}(\tau, \tau')$ and taking the Fourier transform with respect to $\tau -\tau'$, 
we obtain an expression of $G^{\rm c}_{\mib{k}\alpha \mib{k}'\alpha'}({\rm i}\epsilon_n)$.
For impurity models, the site-diagonal component, $G^{\rm c}_{\alpha}({\rm i}\epsilon_n)$, is written in terms of the impurity $t$-matrix $t_{\alpha}({\rm i}\epsilon_n)$ as
\begin{align}
	G^{\rm c}_{\alpha}({\rm i}\epsilon_n) &= g_{\alpha}({\rm i}\epsilon_n)
	+g_{\alpha}({\rm i}\epsilon_n) t_{\alpha}({\rm i}\epsilon_n) g_{\alpha}({\rm i}\epsilon_n), \\
	t_{\alpha}(\tau, \tau') &= -\langle T_{\tau} j_{\alpha}(\tau) j_{\alpha}^{\dag}(\tau') \rangle
	+ \delta(\tau-\tau') \langle \{ j_{\alpha}, c_{\alpha}^{\dag} \} \rangle.
\label{eq:t_matrix_imp}
\end{align}
Its derivation is given in Appendix\ref{sec:deriv_gen_t}.
Figure~\ref{fig:single_particle} shows the corresponding diagrammatical representation. 
\begin{figure}[t]
	\begin{center}
	\includegraphics[width=7cm]{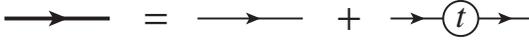}
	\end{center}
	\caption{Diagrammatical representation of the equation for $G^{\rm c}({\rm i}\epsilon_n)$. Thick and thin lines denote the full and bare Green functions, $G^{\rm c}({\rm i}\epsilon_n)$ and $g({\rm i}\epsilon_n)$, respectively.}
	\label{fig:single_particle}
\end{figure}
The impurity $t$-matrix includes all consequences of impurity scattering.
The argument of $t_{\alpha}(\epsilon+{\rm i}0)=|t_{\alpha}(\epsilon)| {\rm e}^{{\rm i}\phi(\epsilon)}$ gives the phase shift of the conduction electrons (see for example ref.~\citen{Yosida-Book}).
In the impurity Anderson model, the second term vanishes and the $t$-matrix is given by the $f$-electron Green function, $t_{\alpha}({\rm i}\epsilon_n) = V^2 G_{\alpha}^f({\rm i}\epsilon_n)$. 
In the impurity CS model, on the other hand, the second term gives the contribution in the first Born approximation $J \langle f_{\alpha}^{\dag}f_{\alpha} \rangle$, which is independent of energy in proportion to an occupation number of the state $\alpha$. 

For the periodic model, the Green function becomes diagonal in the momentum space, $G^{\rm c}_{\mib{k}\alpha}({\rm i}\epsilon_n)$, and is given by
\begin{align}
	G^{\rm c}_{\mib{k}\alpha}({\rm i}\epsilon_n) &= g_{\mib{k}\alpha}({\rm i}\epsilon_n)
	+g_{\mib{k}\alpha}({\rm i}\epsilon_n) t_{\mib{k}\alpha}({\rm i}\epsilon_n) g_{\mib{k}\alpha}({\rm i}\epsilon_n), \\
	t_{\mib{k}\alpha}(\tau, \tau') &= -\langle T_{\tau} j_{\mib{k}\alpha}(\tau) j_{\mib{k}\alpha}^{\dag}(\tau') \rangle
	+ \delta(\tau-\tau') \langle \{ j_{\mib{k}\alpha}, c_{\mib{k}\alpha}^{\dag} \} \rangle.
\label{eq:t_matrix_perio}
\end{align}
In the Anderson lattice model, the $t$-matrix corresponds to $t_{\mib{k}\alpha}({\rm i}\epsilon_n) = V^2 G^f_{\mib{k}\alpha}({\rm i}\epsilon_n)$ as in the impurity model.
In the CS lattice model, the second term gives $(J/N_0) \sum_{\mib{k}'} \langle f_{\mib{k}'\alpha}^{\dag} f_{\mib{k}'\alpha} \rangle$. 
By means of the equations of motion, an expression of the internal energy can be derived from the single-particle Green function (see Appendix\ref{sec:internal_energy}). 

Finally, we conclude by noting the relation between the $t$-matrix and the occupation number of the local states. 
Since the $t$-matrix in the Anderson model is identical to the $f$-electron Green function with the factor $V^2$, 
the $f$-electron number for each component, $n^f_{\alpha}$, is given by
\begin{align}
	T \sum_n t_{\alpha} ({\rm i}\epsilon_n) {\rm e}^{{\rm i}\epsilon_n \delta} = V^2 n^f_{\alpha}, \qquad \text{(Anderson model)}
\label{eq:n_f_AL}
\end{align}
where $\delta$ is a positive infinitesimal. 
In the CS model, on the other hand, the Born term refers to the occupation number. 
Since the first term in eqs.~(\ref{eq:t_matrix_imp}) and (\ref{eq:t_matrix_perio}) vanish in the high-frequency limit, the Born term can be extracted by
\begin{align}
	\lim_{\epsilon_n \rightarrow \infty} t_{\alpha} ({\rm i}\epsilon_n) = J n^f_{\alpha}. \qquad \text{(CS model)}
\label{eq:n_f_CS}
\end{align}
These ways to derive the occupation number from the $t$-matrix will be extended to the two-particle Green function in evaluating the local susceptibility.

\subsection{Two-particle Green function}

By writing down the equations of motion for the single-particle Green function, we have elicited information on the local moment.
Here we extend this procedure to the two-particle correlation functions.
We define generalized susceptibility for conduction electrons with four time variables as follows:
\begin{align}
	\chi_{1234}^{\rm c} (\tau_1, \tau_2&, \tau_3, \tau_4)
	= \langle T_{\tau} c_1^{\dag}(\tau_1) c_2(\tau_2) c_3^{\dag}(\tau_3) c_4(\tau_4) \rangle \nonumber \\
	&- \langle T_{\tau} c_1^{\dag}(\tau_1) c_2(\tau_2) \rangle  \langle T_{\tau} c_3^{\dag}(\tau_3) c_4(\tau_4) \rangle, 
\label{eq:def_chi_c_1234}
\end{align}
where the indices symbolically represent the wavenumber and the spin component. 
The Fourier transform is defined by
\begin{align}
	\chi_{1234}^{\rm c} &({\rm i}\epsilon_n, {\rm i}\epsilon_{n'}; {\rm i}\nu_m) \nonumber \\
	&= \frac{1}{\beta^2} \int_0^{\beta} {\rm d}\tau_1 \cdots \int_0^{\beta} {\rm d}\tau_4
	\chi_{1234}^{\rm c} (\tau_1, \tau_2, \tau_3, \tau_4) \nonumber \\
	&\quad \times {\rm e}^{{\rm i}\epsilon_n (\tau_2-\tau_1)}
	{\rm e}^{{\rm i}\epsilon_{n'} (\tau_4-\tau_3)}
	{\rm e}^{{\rm i}\nu_m (\tau_2-\tau_3)},
\label{eq:Fourier_chi_c_1234}
\end{align}
where $\nu_m=2m\pi T$ is the Matsubara frequency for bosons.
The susceptibility is obtained by letting $\tau_1 = \tau_2$ and $\tau_3 = \tau_4$ in eq.~(\ref{eq:def_chi_c_1234}), so that its Fourier transform $\chi_{1234}^{\rm c}({\rm i}\nu_m)$ is evaluated by
\begin{align}
	\chi_{1234}^{\rm c}({\rm i}\nu_m) = T\sum_{nn'} \chi_{1234}^{\rm c} ({\rm i}\epsilon_n, {\rm i}\epsilon_{n'}; {\rm i}\nu_m).
\end{align}

\begin{figure}[t]
	\begin{center}
	\includegraphics[width=8.5cm]{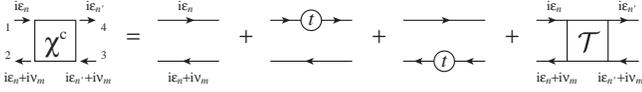}
	\end{center}
	\caption{Diagrammatical representation of the equation for $\chi^{\rm c}_{1234}({\rm i}\epsilon_n, {\rm i}\epsilon_{n'}; {\rm i}\nu_m)$ in eq.~(\ref{eq:chi_c_1234_gen_t}). }
	\label{fig:chi_c_1234}
\end{figure}

By analogy with the single-particle Green function, we consider extracting information of the localized electrons from $\chi^{\rm c}$. 
For this purpose, we define a generalized $t$-matrix $\mathcal{T}_{1234} ({\rm i}\epsilon_n, {\rm i}\epsilon_{n'}; {\rm i}\nu_m)$ for the two-particle Green function as follows:
\begin{align}
	&\chi_{1234}^{\rm c} ({\rm i}\epsilon_n, {\rm i}\epsilon_{n'}; {\rm i}\nu_m) \nonumber \\
	&= - \delta_{nn'} [ \delta_{14}\delta_{23} g_1({\rm i}\epsilon_n) g_2({\rm i}\epsilon_n +{\rm i}\nu_m) \nonumber \\
	 &\quad + \delta_{14} g_1({\rm i}\epsilon_n) \cdot g_3({\rm i}\epsilon_n +{\rm i}\nu_m) t_{32}({\rm i}\epsilon_n +{\rm i}\nu_m) g_2({\rm i}\epsilon_n +{\rm i}\nu_m) \nonumber \\
	 &\quad + g_4({\rm i}\epsilon_n) t_{41}({\rm i}\epsilon_n) g_1({\rm i}\epsilon_n) \cdot \delta_{23} g_2({\rm i}\epsilon_n +{\rm i}\nu_m) ] \nonumber \\
	&+ g_1({\rm i}\epsilon_n) g_2({\rm i}\epsilon_n +{\rm i}\nu_m) g_3({\rm i}\epsilon_{n'} +{\rm i}\nu_m) g_4({\rm i}\epsilon_{n'})
	\nonumber \\
	&\quad \times \mathcal{T}_{1234} ({\rm i}\epsilon_n, {\rm i}\epsilon_{n'}; {\rm i}\nu_m), 
\label{eq:chi_c_1234_gen_t}
\end{align}
which is diagrammatically expressed in Fig.~\ref{fig:chi_c_1234}. 
The generalized $t$-matrix includes all effective interactions between two conduction electrons via the localized electron. 
From an equation of motion for $\chi^{\rm c}$, we can obtain an explicit expression of the generalized $t$-matrix (see Appendix\ref{sec:deriv_gen_t} for detail). 
In the imaginary-time domain, $\mathcal{T}_{1234} ({\rm i}\epsilon_n, {\rm i}\epsilon_{n'}; {\rm i}\nu_m)$ is given by
\begin{align}
	&\mathcal{T}_{1234}(\tau_1, \tau_2, \tau_3, \tau_4) \nonumber \\
	&= \langle T_{\tau} j_1^{\dag}(\tau_1) j_2(\tau_2) j_3^{\dag}(\tau_3) j_4(\tau_4) \rangle 
	 -t_{21}(\tau_2, \tau_1) t_{43}(\tau_4, \tau_3) \nonumber \\
	&+ \delta(\tau_1-\tau_2) \delta(\tau_3-\tau_4) \langle T_{\tau} \{ j_1^{\dag}(\tau_1), c_2(\tau_1) \}
	 \{ j_3^{\dag}(\tau_3), c_4(\tau_3) \} \rangle \nonumber \\
	&- \delta(\tau_1-\tau_4) \delta(\tau_2-\tau_3) \langle T_{\tau} \{ j_1^{\dag}(\tau_1), c_4(\tau_1) \}
	 \{ j_3^{\dag}(\tau_3), c_2(\tau_3) \} \rangle \nonumber \\
	&+ \delta(\tau_1-\tau_2) \langle T_{\tau} \{ j_1^{\dag}(\tau_1), c_2(\tau_1) \} j_3^{\dag}(\tau_3) j_4(\tau_4) \rangle \nonumber \\
	&+ \delta(\tau_3-\tau_4) \langle T_{\tau} j_1^{\dag}(\tau_1) j_2(\tau_2) \{ j_3^{\dag}(\tau_3), c_4(\tau_3) \} \rangle \nonumber \\
	&- \delta(\tau_1-\tau_4) \langle T_{\tau} \{ j_1^{\dag}(\tau_1), c_4(\tau_1) \} j_3^{\dag}(\tau_3) j_2(\tau_2) \rangle \nonumber \\
	&- \delta(\tau_2-\tau_3) \langle T_{\tau} j_1^{\dag}(\tau_1) j_4(\tau_4) \{ j_3^{\dag}(\tau_3), c_2(\tau_3) \} \rangle.
\label{eq:def_gen_t}
\end{align}
In this derivation, we have used the relations $\{ j_1^{\dag}, c_2 \} = \{c_1^{\dag}, j_2 \}$ and $[ c_1, \{j_2^{\dag}, c_3\} ] = [ c_1^{\dag}, \{j_2^{\dag}, c_3\} ] =0$.
The latter relation holds in the case where $H_{\rm loc}$ is bilinear in $c^{\dag}$ and $c$. 

\begin{figure}[t]
	\begin{center}
	\includegraphics[width=8.5cm]{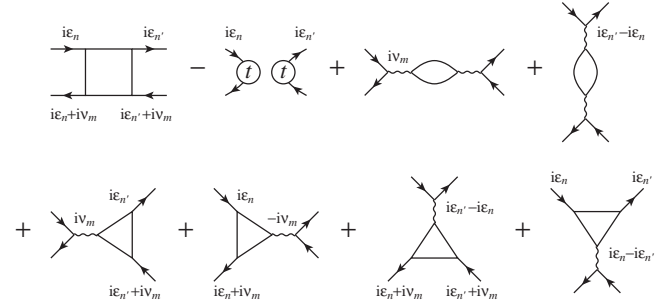}
	\end{center}
	\caption{Diagrammatical representations of terms in the generalized $t$-matrix, eq.~(\ref{eq:def_gen_t}). Wavy lines indicate the delta-function for a time variable. }
	\label{fig:gen_t_term}
\end{figure}

In the Anderson model, terms except for the first two vanish in eq.~(\ref{eq:def_gen_t}) because of the relation $\{ j_1^{\dag}, c_2 \}=0$, which follows from eq.~(\ref{eq:def_j}).
Hence $\mathcal{T}$ corresponds to the two-particle Green function of the localized electron as 
$\mathcal{T}_{1234}({\rm i}\epsilon_n, {\rm i}\epsilon_{n'}; {\rm i}\nu_m)=V^4 \chi^f_{1234}({\rm i}\epsilon_n, {\rm i}\epsilon_{n'}; {\rm i}\nu_m)$, where $\chi^f_{1234}$ is defined in a manner similar to $\chi^{\rm c}_{1234}$.
Therefore, in the Anderson model, the susceptibility $\chi^f({\rm i}\nu_m)$ is evaluated by
\begin{align}
	\chi^f_{1234}({\rm i}\nu_m) = \frac{T}{V^4} \sum_{nn'} \mathcal{T}_{1234}({\rm i}\epsilon_n, {\rm i}\epsilon_{n'}; {\rm i}\nu_m). \nonumber \\
	\text{(Anderson model)}
\label{eq:chi_f_1234_Anderson}
\end{align}
In the CS model, on the other hand, all terms having the delta-functions contribute to the generalized $t$-matrix in eq.~(\ref{eq:def_gen_t}).
Figure~\ref{fig:gen_t_term} shows the corresponding diagrams in the frequency domain. 
Correlation functions vanish in the high-frequency limit, while the Fourier transform of $\delta(\tau)$ is constant in the frequency domain.
Thus it is possible to extract each term in eq.~(\ref{eq:def_gen_t}) by taking a limit for the corresponding variables. 
For example, the third term, which has $\delta(\tau_1-\tau_2) \delta(\tau_3-\tau_4)$, can be picked out by a limit with respect to $\epsilon_n$ and $\epsilon_{n'}$ such as
\begin{align}
	&\lim_{\epsilon_n \rightarrow +\infty} \lim_{\epsilon_{n'} \rightarrow -\infty}
	\mathcal{T}_{1234} ({\rm i}\epsilon_n, {\rm i}\epsilon_{n'}; {\rm i}\nu_m) \nonumber \\
	&= \frac{J^2}{\beta^2} \int_0^{\beta} {\rm d}\tau_2 \int_0^{\beta} {\rm d}\tau_3
	 \Big[ \langle T_{\tau} f_1^{\dag}(\tau_2) f_2(\tau_2) f_3^{\dag}(\tau_3) f_4(\tau_3) \rangle \nonumber \\
	 &\quad - \langle f_1^{\dag} f_2 \rangle  \langle f_3^{\dag} f_4 \rangle \Big]
	 {\rm e}^{{\rm i}\nu_m (\tau_2-\tau_3)} \nonumber \\
	&= T J^2 \chi^f_{1234}({\rm i}\nu_m), 
	\qquad \text{(CS model)}
\label{eq:chi_f_1234_CS}
\end{align}
which is nothing but the susceptibility of the localized electron. 
The second term in the middle row comes from the second term in eq.~(\ref{eq:def_gen_t}), whose high-frequency limit is given in eq.~(\ref{eq:n_f_CS}). 
We note that, in eq.~(\ref{eq:chi_f_1234_CS}), the signs of $\epsilon_n$ and $\epsilon_{n'}$ should be different from each other.
If we take another limit keeping the difference $\nu_{m'}=\epsilon_{n'}-\epsilon_n$, the fourth term in eq.~(\ref{eq:def_gen_t}) remains, so that we obtain
\begin{align}
	&\lim_{\epsilon_n \rightarrow +\infty} \lim_{\epsilon_{n'} \rightarrow +\infty}
	\mathcal{T}_{1234} ({\rm i}\epsilon_n, {\rm i}\epsilon_{n'}; {\rm i}\nu_m) \nonumber \\
	&= T J^2 \left[ \chi^f_{1234}({\rm i}\nu_m) - \chi^f_{1432}({\rm i}\nu_{m'}) - \langle f^{\dag}_1 f_4 \rangle \langle f^{\dag}_3 f_2 \rangle \right]. \nonumber \\
	&\hspace{5cm} \text{(CS model)}
\end{align}

In conclusion, in order to obtain the susceptibility of the localized electrons, we may evaluate the generalized $t$-matrix $\mathcal{T}$. 
Eliminating the fermion frequencies by certain ways, $\mathcal{T}$ yields the two-particle correlations of the localized electrons. 
The ways to obtain the susceptibilities are different in the Anderson and the CS models.
This difference originates in the relations between the occupation number and the $t$-matrix, eqs.~(\ref{eq:n_f_AL}) and (\ref{eq:n_f_CS}).

We set up an integral equation for $\mathcal{T}_{1234}({\rm i}\epsilon_n, {\rm i}\epsilon_{n'}; {\rm i}\nu_m)$. 
It has been shown in eq.~(\ref{eq:chi_f_1234_Anderson}) for the Anderson model that $\mathcal{T}_{1234}$ corresponds to the generalized susceptibility of $f$ electrons, $\chi_{1234}^f({\rm i}\epsilon_n, {\rm i}\epsilon_{n'}; {\rm i}\nu_m)$, where the irreducible vertex part is defined by the Bethe-Salpeter equation.
In a similar manner, we introduce an irreducible vertex part of $\mathcal{T}_{1234}$ by
\begin{align}
	&\mathcal{T}_{1234} ({\rm i}\epsilon_n, {\rm i}\epsilon_{n'}; {\rm i}\nu_m)
	= \delta_{nn'} \mathcal{T}^0_{1234} ({\rm i}\epsilon_n; {\rm i}\nu_m) \nonumber \\
	&+ T \sum_{n''} \sum_{1'2'3'4'} \mathcal{T}^0_{122'1'} ({\rm i}\epsilon_n; {\rm i}\nu_m)
	 I_{1'2'3'4'}({\rm i}\epsilon_n, {\rm i}\epsilon_{n''}; {\rm i}\nu_m) \nonumber \\
	&\qquad \times \mathcal{T}_{4'3'34} ({\rm i}\epsilon_{n''}, {\rm i}\epsilon_{n'}; {\rm i}\nu_m),
\label{eq:T_1234_BS}
\end{align}
\begin{figure}[t]
	\begin{center}
	\includegraphics[width=8.5cm]{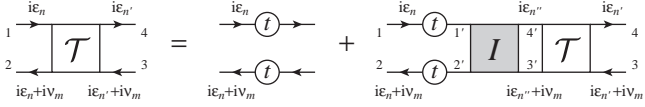}
	\end{center}
	\caption{An integral equation for the generalized $t$-matrix $\mathcal{T}({\rm i}\epsilon_n, {\rm i}\epsilon_{n'}; {\rm i}\nu_m)$. }
	\label{fig:gen_t_eq_1234}
\end{figure}
where $\mathcal{T}^0_{1234}({\rm i}\epsilon_n; {\rm i}\nu_m)$ is defined by
\begin{align}
	\mathcal{T}^0_{1234} ({\rm i}\epsilon_n; {\rm i}\nu_m) = - t_{41}({\rm i}\epsilon_n) t_{23}({\rm i}\epsilon_n +{\rm i}\nu_m).
\end{align}
The above equation is diagrammatically represented in Fig.~\ref{fig:gen_t_eq_1234}. 
In the Anderson model, $\mathcal{T}^0$ and $I$ correspond to $V^4 \chi^{f0}$ and $\Gamma^f /V^4$, respectively, with $\Gamma^f$ being the irreducible vertex part.

With these formulae, we now describe the following expression of the susceptibility in the lattice:
\begin{align}
	&\chi^{\rm c}_{\mib{k} \mib{k}' \mib{q}, \alpha \alpha'} (\tau_1, \tau_2, \tau_3, \tau_4) \nonumber \\
	&= \langle T_{\tau} c_{\mib{k},\alpha}^{\dag}(\tau_1) c_{\mib{k}+\mib{q},\alpha}(\tau_2)
	 c_{\mib{k}'+\mib{q}, \alpha'}^{\dag}(\tau_3) c_{\mib{k}', \alpha'}(\tau_4) \rangle \nonumber \\
	&- \delta_{\mib{q},0} \langle T_{\tau} c_{\mib{k},\alpha}^{\dag}(\tau_1) c_{\mib{k},\alpha}(\tau_2) \rangle 
	 \langle T_{\tau} c_{\mib{k}', \alpha'}^{\dag}(\tau_3) c_{\mib{k}', \alpha'}(\tau_4) \rangle, 
\label{eq:def_chi_c}
\end{align}
The Fourier transform, defined in eq.~(\ref{eq:Fourier_chi_c_1234}), leads to
$\chi^{\rm c}_{\mib{k} \mib{k}' \mib{q}, \alpha \alpha'} ({\rm i}\epsilon_n, {\rm i}\epsilon_{n'}; {\rm i}\nu_m)$.
To discuss physical responses to external fields, we introduce two response functions constructed by linear combinations of spin components.
One is the charge response function $\chi^{\rm c, chg}$ defined from the charge density operator $\sum_{\alpha} n^{\rm c}_{\alpha}$, and 
another is the magnetic response function $\chi^{\rm c, mag}$ defined from the magnetic moment, $\sum_{\alpha} m_{\alpha} n^{\rm c}_{\alpha}$ with $\sum_{\alpha} m_{\alpha}=0$: 
\begin{align}
	\chi^{\rm c, chg}_{\mib{k} \mib{k}' \mib{q}} ({\rm i}\epsilon_n, {\rm i}\epsilon_{n'}; {\rm i}\nu_m)
	&= \sum_{\alpha \alpha'} \chi^{\rm c}_{\mib{k} \mib{k}' \mib{q}, \alpha \alpha'}({\rm i}\epsilon_n, {\rm i}\epsilon_{n'}; {\rm i}\nu_m), \\
	\chi^{\rm c, mag}_{\mib{k} \mib{k}' \mib{q}} ({\rm i}\epsilon_n, {\rm i}\epsilon_{n'}; {\rm i}\nu_m)
	&= \sum_{\alpha \alpha'} m_{\alpha} m_{\alpha'} \chi^{\rm c}_{\mib{k} \mib{k}' \mib{q}, \alpha \alpha'}({\rm i}\epsilon_n, {\rm i}\epsilon_{n'}; {\rm i}\nu_m).
\end{align}
We assume the SU($N$) symmetry. 
Then the above quantities may be rewritten as 
\begin{align}
	\chi^{\rm c, chg}_{\mib{k} \mib{k}' \mib{q}} &= N \chi^{\rm c, diag}_{\mib{k} \mib{k}' \mib{q}} + N(N-1) \chi^{\rm c, offd}_{\mib{k} \mib{k}' \mib{q}}, \\
	\chi^{\rm c, mag}_{\mib{k} \mib{k}' \mib{q}} &= N C_N ( \chi^{\rm c, diag}_{\mib{k} \mib{k}' \mib{q}} - \chi^{\rm c, offd}_{\mib{k} \mib{k}' \mib{q}} ),
\end{align}
where $\chi^{\rm c, diag}$ and $\chi^{\rm c, offd}$ denote respectively the diagonal and off-diagonal element of $\chi^{\rm c}_{\alpha \alpha'}$ in spin indices, 
and $C_N=N^{-1}\sum_{\alpha} m_{\alpha}^2$ is the Curie constant. 
The physical response is evaluated from the generalized susceptibility by
\begin{align}
	\chi_{\mib{q}}^{{\rm c}, \gamma} ({\rm i}\nu_m)
	= \frac{1}{N_0} \sum_{\mib{k}\mib{k}'} T\sum_{nn'} \chi_{\mib{k}\mib{k}'\mib{q}}^{{\rm c}, \gamma} ({\rm i}\epsilon_n, {\rm i}\epsilon_{n'}; {\rm i}\nu_m),
\end{align}
where $\gamma=\text{chg, mag}$ represents the charge and magnetic channels.

We introduce the generalized $t$-matrix $\mathcal{T}^{\gamma}_{\mib{k} \mib{k}' \mib{q}} ({\rm i}\epsilon_n, {\rm i}\epsilon_{n'}; {\rm i}\nu_m)$ for each channel $\gamma$ to describe the response functions of the localized moments. 
From eq.~(\ref{eq:chi_c_1234_gen_t}), $\chi^{{\rm c}, \gamma}_{\mib{k} \mib{k}' \mib{q}}$ gives $\mathcal{T}^{\gamma}_{\mib{k} \mib{k}' \mib{q}}$ as follows:
\begin{align}
	&\chi^{{\rm c}, \gamma}_{\mib{k} \mib{k}' \mib{q}} ({\rm i}\epsilon_n, {\rm i}\epsilon_{n'}; {\rm i}\nu_m) \nonumber \\
	&= - N C_N^{\gamma} \delta_{nn'} \delta_{\mib{k} \mib{k}'}[ g(k) g(k+q) \nonumber \\
	 &\qquad + g(k) \cdot g(k+q)^2 t(k+q)
	 + g(k)^2 t(k) \cdot g(k+q) ] \nonumber \\
	&+ g(k) g(k+q) g(k'+q) g(k')
	 \mathcal{T}^{\gamma}_{\mib{k} \mib{k}' \mib{q}} ({\rm i}\epsilon_n, {\rm i}\epsilon_{n'}; {\rm i}\nu_m), 
\label{eq:chi_c_gen_t}
\end{align}
where four-dimensional vectors are introduced by $k=(\mib{k}, {\rm i}\epsilon_n)$, $k'=(\mib{k}', {\rm i}\epsilon_{n'})$ and $q=(\mib{q}, {\rm i}\nu_m)$. 
The factor $C_N^{\gamma}$ corresponds to $C_N$ for the magnetic channel and to unity for the charge channel.
%
%
In the Anderson model, eq.~(\ref{eq:chi_f_1234_Anderson}) yields the susceptibility of the localized electrons, so that we obtain
\begin{align}
	\chi^{f, \gamma}_{\mib{q}} ({\rm i}\nu_m)
	= \frac{T}{V^4} \sum_{nn'} &\mathcal{T}^{\gamma}_{\mib{q}} ({\rm i}\epsilon_n, {\rm i}\epsilon_{n'}; {\rm i}\nu_m),\nonumber \\
	&\qquad \text{(Anderson model)}
\label{eq:T_q_sum}
\end{align}
where $\mathcal{T}^{\gamma}_{\mib{q}}$ is defined by $N_0^{-1} \sum_{\mib{k}\mib{k}'} \mathcal{T}^{\gamma}_{\mib{k}\mib{k}'\mib{q}}$.
Here, we have assumed that the matrix elements of the magnetic moment of the localized electron are the same as those of conduction electrons.
In the CS model, on the other hand, from eq.~(\ref{eq:chi_f_1234_CS}), $\mathcal{T}^{\gamma}_{\mib{q}}$ gives $\chi_{\mib{q}}^{f, \gamma}({\rm i}\nu_m)$ in the limit
\begin{align}
	\chi^{f, \gamma}_{\mib{q}} ({\rm i}\nu_m)
	= \frac{1}{T J^2} \lim_{\epsilon_n \rightarrow +\infty} \lim_{\epsilon_{n'} \rightarrow -\infty}
	&\mathcal{T}^{\gamma}_{\mib{q}} ({\rm i}\epsilon_n, {\rm i}\epsilon_{n'}; {\rm i}\nu_m). \nonumber \\
	& \qquad \text{(CS model)}
\label{eq:T_q_limit}
\end{align}
By definition, the magnetic channel is zero in the CS model, so that we represent $\chi^{f, {\rm mag}}_{\mib{q}}$ simply by $\chi^f_{\mib{q}}$ hereafter.

\section{Dynamical Mean-Field Theory for Localized Models}
\label{sec:dmft}

In infinite dimensions, the self-energy part due to local interactions is also local.
Accordingly, the periodic system with local interactions can be reduced to a single-impurity model in an effective medium. 
The DMFT determines the effective medium from the local self-energy in a self-consistent manner\cite{Georges, Kuramoto-Kitaoka}.
Spatial dependences of two-particle correlation functions in the periodic system can be evaluated from the local vertices in the effective impurity system\cite{Georges, Jarrell-AL, Jarrell-Hubbard}.

In this section, we apply the DMFT formalism to the CS lattice, where the ordinary Green function cannot be defined for localized electrons.
For this application, the $t$-matrix and its two-particle generalization, introduced in the previous section, play a key role.
The self-consistent equation for the conduction electrons will be constructed with the $t$-matrix.
Later in this section, we further derive spatial dependences of two-particle correlation functions with use of the generalized $t$-matrix.

\subsection{Self-consistent equations}
It is convenient to begin with the Anderson lattice model, since its self-consistent equation has been well established. 
In the DMFT, an effect of $U$ at the surrounding sites is incorporated in an unperturbed Green function. 
It is referred to as a cavity Green function $\mathcal{G}^{f0}(z)$.
Letting $\mathcal{G}^f(z)$ be the full Green function in the effective impurity system, the self-energy part $\Sigma^f(z)$ is given by
\begin{align}
	\Sigma^f(z) = \mathcal{G}^{f0} (z)^{-1} - \mathcal{G}^f(z)^{-1}.
\label{eq:self_f}
\end{align}
In the infinite-dimensional limit, the self-energy part becomes local, so that $\Sigma^f(z)$ is equivalent to that of the periodic system. 
The site-diagonal Green function $\bar{G}^f(z)$ of the periodic system is given in terms of $\Sigma^f(z)$ by
\begin{align}
	\bar{G}^f(z) = \int {\rm d}\epsilon \rho(\epsilon)
	 \left[ z - \epsilon_f - \Sigma^f(z) - \frac{V^2}{z -\epsilon + \mu} \right]^{-1}, 
\label{eq:G_f_site-diag}
\end{align}
where $\rho(\epsilon) = N_0^{-1} \sum_{\mib{k}} \delta(\epsilon - \epsilon_{\mib{k}})$ is the density of states of conduction electrons. 
The self-consistency condition $\mathcal{G}^f(z)=\bar{G}^f(z)$ renews cavity Green function as
\begin{align}
	\mathcal{G}^{f0}(z)^{-1} = \bar{G}^f(z)^{-1} + \Sigma^f(z). 
\label{eq:cav_f}
\end{align}
This equation corresponds to a procedure which removes $\Sigma^f(z)$ at one site keeping $\Sigma^f(z)$ at the surrounding sites. 
A self-consistent solution of the above equations yields the Green function of the lattice in the DMFT. 

In order to employ a perturbation theory from the atomic limit as an impurity solver, it is convenient to rewrite the above equations in terms of the Green function of conduction electrons.
This transform enables us to apply the DMFT formalism to the localized model, such as the Kondo lattice and the CS lattice. 
To this end, we introduce a cavity field $\mathcal{G}^{\rm c0}(z)$ for conduction electrons defined by $\mathcal{G}^{f0}(z) = [z - \epsilon_f -V^2 \mathcal{G}^{\rm c0}(z)]^{-1}$. 
The self-energy part is local in the infinite-dimensional limit, so that the site-diagonal element $\bar{G}^{\rm c}(z)$ of the conduction-electron Green function is given by 
\begin{align}
	\bar{G}^{\rm c}(z) = \int {\rm d}\epsilon \rho(\epsilon)
	 \left[ z - \epsilon + \mu - \Sigma^{\rm c}(z) \right]^{-1}.
\label{eq:G_c}
\end{align}
The self-energy part $\Sigma^{\rm c}(z)$ is defined by
\begin{align}
	\Sigma^{\rm c}(z) = \frac{V^2}{z - \epsilon_f - \Sigma^f(z)}
	= \frac{t(z)}{1+\mathcal{G}^{\rm c0}(z) t(z)},
\label{eq:self_c}
\end{align}
where the impurity $t$-matrix $t(z)$ is defined by $t(z)=V^2 \mathcal{G}^{f}(z)$, and is related to the site-diagonal Green function $\mathcal{G}^{\rm c}(z)$ in the effective impurity system by
\begin{align}
	\mathcal{G}^{\rm c} (z) = \mathcal{G}^{\rm c0}(z) + \mathcal{G}^{\rm c0}(z) t(z) \mathcal{G}^{\rm c0}(z).
\label{eq:Gc_eff_imp}
\end{align}
It is obvious from eq.~(\ref{eq:self_c}) that $\Sigma^{\rm c}(z)$ characterizes an impurity scattering without repetition, while the $t$-matrix incorporates all scattering processes including repetitions of $\Sigma^{\rm c}(z)$. 
The self-consistency condition for the localized electrons, $\mathcal{G}^f(z) = \bar{G}^f(z)$, ensures a similar relation for the conduction electrons, $\mathcal{G}^{\rm c}(z) = \bar{G}^{\rm c}(z)$. 
Hence the cavity field $\mathcal{G}^{\rm c0}(z)$ is given from $\bar{G}^{\rm c}(z)$ as
\begin{align}
	\mathcal{G}^{\rm c0}(z)^{-1} &= \bar{G}^{\rm c}(z)^{-1} + \Sigma^{\rm c}(z).
\label{eq:cav_c}
\end{align}
Equations~(\ref{eq:G_c})--(\ref{eq:cav_c}), written in terms of conduction electrons and the $t$-matrix, are identical with eqs.~(\ref{eq:self_f})--(\ref{eq:cav_f}), written in terms of the localized electrons. 
The DMFT equation for the CS lattice model
eventually becomes the same as that of 
the Hubbard model. 
The sources of the self-energy $\Sigma^{\rm c}$ are, however, different: 
in the CS lattice model, $\Sigma^{\rm c}$ signifies the impurity scattering and accordingly, is obtained from the impurity $t$-matrix in eq.~(\ref{eq:self_c}), while in the Hubbard model the self-energy arises from the Coulomb interaction between conduction electrons. 
From eq.~(\ref{eq:G_f_site-diag}), the $t$-matrix of the lattice, $\bar{t}(z)$, defined by $\bar{t}(z) = V^2 \bar{G}^f(z)$ is related to $\bar{G}^{\rm c}(z)$ as
\begin{align}
	\bar{t}(z) = \Sigma^{\rm c}(z) + \Sigma^{\rm c}(z)^2 \bar{G}^{\rm c}(z). 
\label{eq:t_site-diag}
\end{align}
The impurity $t$-matrix $t(z)$ equals to $\bar{t}(z)$ from the self-consistency condition.
In the localized model, it is possible to study formation of heavy quasi-particles through the $t$-matrix of the lattice.

The DMFT imposes self-consistency on the site-diagonal element of the Green function, so that only $\mib{k}$-averaged quantities appear in the equations. 
However, it is possible to see the $\mib{k}$-dependence through the integrand of the average. 
The Green function and the $t$-matrix for a given $\mib{k}$ are evaluated from $\Sigma^{\rm c}(z)$ by
\begin{align}
	G^{\rm c}_{\mib{k}} (z) &= [z-\epsilon_{\mib{k}} + \mu -\Sigma^{\rm c}(z)]^{-1}
	= \frac{g_{\mib{k}}(z)}{1-g_{\mib{k}}(z) \Sigma^{\rm c}(z)},
\label{eq:Gc_k} \\
	t_{\mib{k}}(z) &= \Sigma^{\rm c}(z) + \Sigma^{\rm c}(z)^2 G^{\rm c}_{\mib{k}}(z)
	= \frac{\Sigma^{\rm c}(z)}{1-g_{\mib{k}}(z) \Sigma^{\rm c}(z)},
\label{eq:t_k}
\end{align}
where the $\mib{k}$-dependence enters only through the energy $\epsilon_{\mib{k}}$ of the conduction electrons.

\subsection{Spatial dependence of two-particle correlations}

Let us consider the two-particle Green function within the DMFT.
Applying eq.~(\ref{eq:chi_c_1234_gen_t}) to the effective impurity system, the local component of the generalized $t$-matrix $\mathcal{T}^{\gamma}_{\rm loc}$ is defined by
\begin{align}
	&\chi^{{\rm c}, \gamma}_{\rm loc} ({\rm i}\epsilon_n, {\rm i}\epsilon_{n'}; {\rm i}\nu_m) \nonumber \\
	&= - N C_N^{\gamma} \delta_{nn'} [ \mathcal{G}^{\rm c0}({\rm i}\epsilon_n) \mathcal{G}^{\rm c0}({\rm i}\epsilon_n +{\rm i}\nu_m) \nonumber \\
	 &\qquad + \mathcal{G}^{\rm c0}({\rm i}\epsilon_n) \cdot \mathcal{G}^{\rm c0}({\rm i}\epsilon_n +{\rm i}\nu_m)^2 \bar{t}({\rm i}\epsilon_n +{\rm i}\nu_m) \nonumber \\
	 &\qquad + \mathcal{G}^{\rm c0}({\rm i}\epsilon_n)^2 \bar{t}({\rm i}\epsilon_n) \cdot \mathcal{G}^{\rm c0}({\rm i}\epsilon_n +{\rm i}\nu_m) ] \nonumber \\
	&+ \mathcal{G}^{\rm c0}({\rm i}\epsilon_n) \mathcal{G}^{\rm c0}({\rm i}\epsilon_n +{\rm i}\nu_m) \mathcal{G}^{\rm c0}({\rm i}\epsilon_{n'} +{\rm i}\nu_m) \mathcal{G}^{\rm c0}({\rm i}\epsilon_{n'}) \nonumber \\
	&\quad \times \mathcal{T}^{\gamma}_{\rm loc} ({\rm i}\epsilon_n, {\rm i}\epsilon_{n'}; {\rm i}\nu_m).
\end{align}
In a manner similar to eqs.~(\ref{eq:T_q_sum}) and (\ref{eq:T_q_limit}), $\mathcal{T}^{\gamma}_{\rm loc}$ gives the local susceptibilities for each model. 
Noting eq.~(\ref{eq:Gc_eff_imp}), the above equation may be rewritten as 
\begin{align}
	\chi^{{\rm c}, \gamma}_{\rm loc} ({\rm i}\epsilon_n, {\rm i}\epsilon_{n'}; {\rm i}\nu_m)
	&= \delta_{nn'} \chi^{{\rm c0}, \gamma}_{\rm loc} ({\rm i}\epsilon_n; {\rm i}\nu_m) \nonumber \\
	+ \pi ({\rm i}\epsilon_n; {\rm i}\nu_m)
	&[ \mathcal{T}_{\rm loc}^{\gamma} ({\rm i}\epsilon_n, {\rm i}\epsilon_{n'}; {\rm i}\nu_m) \nonumber \\
	 -& \delta_{nn'} \mathcal{T}_{\rm loc}^{0, \gamma} ({\rm i}\epsilon_n; {\rm i}\nu_m) ]
	\pi ({\rm i}\epsilon_{n'}; {\rm i}\nu_m),
\label{eq:chi_c_loc_gen_t}
\end{align}
where we have introduced
\begin{align}
	&\pi ({\rm i}\epsilon_n; {\rm i}\nu_m) = -\mathcal{G}^{\rm c0}({\rm i}\epsilon_n) \mathcal{G}^{\rm c0}({\rm i}\epsilon_n + {\rm i}\nu_m), \nonumber \\
	&\Pi_{\rm loc} ({\rm i}\epsilon_n; {\rm i}\nu_m)
	= -\bar{G}^{\rm c} ({\rm i}\epsilon_n) \bar{G}^{\rm c} ({\rm i}\epsilon_n+{\rm i}\nu_m),
\end{align}
and, $\chi^{{\rm c0}, \gamma}_{\rm loc}$ is defined by $\chi^{{\rm c0}, \gamma}_{\rm loc}=N C_N^{\gamma} \Pi_{\rm loc}$.

We direct our attention to the generalized $t$-matrix in the effective impurity as well as the lattice systems.
We set up an integral equation for $\mathcal{T}^{\gamma}_{\rm loc}$. 
Equation~(\ref{eq:T_1234_BS}) holds for each channel independently, so that we obtain
\begin{align}
	&\mathcal{T}_{\rm loc}^{\gamma} ({\rm i}\epsilon_n, {\rm i}\epsilon_{n'}; {\rm i}\nu_m)
	= \delta_{nn'} \mathcal{T}^{0, \gamma}_{\rm loc} ({\rm i}\epsilon_n; {\rm i}\nu_m) \nonumber \\
	&+ T \sum_{n''} \mathcal{T}^{0, \gamma}_{\rm loc} ({\rm i}\epsilon_n; {\rm i}\nu_m)
	 I^{\gamma}({\rm i}\epsilon_n, {\rm i}\epsilon_{n''}; {\rm i}\nu_m)
	 \mathcal{T}_{\rm loc}^{\gamma} ({\rm i}\epsilon_{n''}, {\rm i}\epsilon_{n'}; {\rm i}\nu_m), 
\label{eq:T_loc}
\end{align}
where $\mathcal{T}^{0, \gamma}_{\rm loc}({\rm i}\epsilon_n; {\rm i}\nu_m)$ is defined by
\begin{align}
	\mathcal{T}^{0, \gamma}_{\rm loc} ({\rm i}\epsilon_n; {\rm i}\nu_m)
	= - N C_N^{\gamma} \bar{t}({\rm i}\epsilon_n) \bar{t}({\rm i}\epsilon_n +{\rm i}\nu_m). 
\label{eq:T0_loc}
\end{align}
In the Anderson model, $I^{\gamma}$ relates to the irreducible vertex part $\Gamma^{f, {\gamma}}$ by $I^{\gamma}= \Gamma^{f, {\gamma}}/V^4$, and $\mathcal{T}^0$ to the generalized susceptibility without the vertex correction, $\chi^{f0, \gamma}$, by $\mathcal{T}^{0, \gamma}=V^4 \chi^{f0, \gamma}$. 
In infinite dimensions, only diagrams for intersite processes are those connected by two propagators.
It follows that $I^{\gamma}$ is a local quantity\cite{Zlatic-Horvatic, Jarrell-Hubbard}.
Therefore connecting $I^{\gamma}$ at each site by a pair of the lattice $t$-matrix leads to the generalized $t$-matrix $\mathcal{T}^{\gamma}_{ij}$ of the periodic systems. 
Taking the Fourier transform over the spatial indices, we obtain an equation for $\mathcal{T}^{\gamma}_{\mib{q}}$ as
\begin{align}
	&\mathcal{T}_{\mib{q}}^{\gamma} ({\rm i}\epsilon_n, {\rm i}\epsilon_{n'}; {\rm i}\nu_m)
	= \delta_{nn'} \mathcal{T}_{\mib{q}}^{0, \gamma} ({\rm i}\epsilon_n; {\rm i}\nu_m) \nonumber \\
	&+ T\sum_{n''} \mathcal{T}_{\mib{q}}^{0, \gamma} ({\rm i}\epsilon_n; {\rm i}\nu_m)
	 I^{\gamma} ({\rm i}\epsilon_n, {\rm i}\epsilon_{n''}; {\rm i}\nu_m)
	 \mathcal{T}_{\mib{q}}^{\gamma} ({\rm i}\epsilon_{n''}, {\rm i}\epsilon_{n'}; {\rm i}\nu_m),
\label{eq:T_q}
\end{align}
where $\mathcal{T}_{\mib{q}}^{0, \gamma}$ is defined by
\begin{align}
	\mathcal{T}_{\mib{q}}^{0, \gamma} ({\rm i}\epsilon_n; {\rm i}\nu_m)
	= - \frac{N C_N^{\gamma}}{N_0} \sum_{\mib{k}} t_{\mib{k}}({\rm i}\epsilon_n) t_{\mib{k}+\mib{q}}({\rm i}\epsilon_n +{\rm i}\nu_m).
\label{eq:T0_q}
\end{align}
We represent eqs.~(\ref{eq:T_loc}) and (\ref{eq:T_q}) in a matrix form for the fermion frequencies as follows:
\begin{align}
	\mathcal{T}_{\rm loc}^{\gamma} &= \mathcal{T}_{\rm loc}^{0, \gamma} + \mathcal{T}_{\rm loc}^{0, \gamma} I^{\gamma} \mathcal{T}_{\rm loc}^{\gamma}, \nonumber \\
	\mathcal{T}_{\mib{q}}^{\gamma} &= \mathcal{T}_{\mib{q}}^{0, \gamma} + \mathcal{T}_{\mib{q}}^{0, \gamma} I^{\gamma} \mathcal{T}_{\mib{q}}^{\gamma}.
\end{align}
Eliminating $I^{\gamma}$ from the above two equations, we obtain the following equation for $\mathcal{T}^{\gamma}_{\mib{q}}$:
\begin{align}
	[\mathcal{T}_{\mib{q}}^{\gamma}]^{-1} = [\mathcal{T}_{\rm loc}^{\gamma}]^{-1} - [\mathcal{T}_{\rm loc}^{0, \gamma}]^{-1} + [\mathcal{T}_{\mib{q}}^{0, \gamma}]^{-1},
\label{eq:T_q_matrix}
\end{align}
which holds for each value of the energy transfer $\nu_m$ independently. 
This equation is an extension of those in the Hubbard model\cite{Jarrell-Hubbard} and the Anderson lattice model\cite{Jarrell-AL} to the CS lattice model.
The corresponding equation for the Anderson lattice model has also been derived within the extended non-crossing approximation (XNCA), where the irreducible vertex is explicitly given in the framework of the NCA\cite{Kuramoto-Watanabe}.
We again note that, in the CS lattice model, a limit of $\mathcal{T}_{\mib{q}}^{\rm mag}$ gives the magnetic susceptibility of the localized moments in eq.~(\ref{eq:T_q_limit}). 
Although eq.~(\ref{eq:T_q_matrix}) is an infinite-sized matrix equation, we actually do not need in numerical calculations to compute $\mathcal{T}$ in the whole range of frequencies. 
The matrix equation as well as the limiting operation in eq.~(\ref{eq:T_q_limit}) can be evaluated with small matrices efficiently (see Appendix\ref{sec:high_freq_limit}).

We return our attention to the susceptibility of conduction electrons. 
The susceptibility makes it possible to discuss instabilities against the charge-density wave (CDW) and spin-density wave (SDW) states, corresponding to the charge and magnetic channel, respectively.
For the generalized $t$-matrix, we have introduced the irreducible vertex $I^{\gamma}$ to derive spatially dependent function. 
To obtain an integral equation for the conduction-electron susceptibility, we substitute eq.~(\ref{eq:T_loc}) into eq.~(\ref{eq:chi_c_loc_gen_t}).
After some manipulations, we obtain the following equation:
\begin{align}
	&\chi^{{\rm c}, \gamma}_{\rm loc} ({\rm i}\epsilon_n, {\rm i}\epsilon_{n'}; {\rm i}\nu_m)
	= \delta_{nn'} \chi^{{\rm c0}, \gamma}_{\rm loc} ({\rm i}\epsilon_n; {\rm i}\nu_m) \nonumber \\
	&+ T\sum_{n''} \delta_{nn'} \chi^{{\rm c0}, \gamma}_{\rm loc} ({\rm i}\epsilon_n; {\rm i}\nu_m)
	\Gamma^{{\rm c}, \gamma} ({\rm i}\epsilon_n, {\rm i}\epsilon_{n''}; {\rm i}\nu_m)
	\chi^{{\rm c}, \gamma}_{\rm loc} ({\rm i}\epsilon_{n''}, {\rm i}\epsilon_{n'}; {\rm i}\nu_m),
\label{eq:chi_c_loc}
\end{align}
For more details and the explicit expression of $\Gamma^{{\rm c}, \gamma}$, see Appendix\ref{sec:vertex_cond}. 
\begin{figure}[t]
	\begin{center}
	\includegraphics[width=8.5cm]{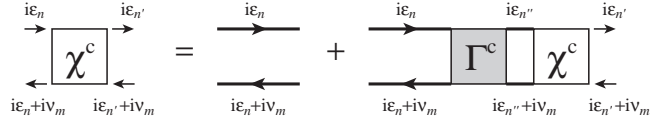}
	\end{center}
	\caption{An integral equation for the generalized susceptibility of conduction electrons, $\chi^{\rm c}({\rm i}\epsilon_n, {\rm i}\epsilon_{n'}; {\rm i}\nu_m)$. }
	\label{fig:chi_c_eq}
\end{figure}
Figure~\ref{fig:chi_c_eq} represents the above equation diagrammatically. 
Since $\Gamma^{{\rm c}, \gamma}$ consists only of local quantities in infinite dimensions, $\Gamma^{{\rm c}, \gamma}$ is also local. 
Hence, with use of the same vertex, $\chi^{{\rm c}, \gamma}_{\mib{q}}$ can be constructed as follows:
\begin{align}
	&\chi^{{\rm c}, \gamma}_{\mib{q}} ({\rm i}\epsilon_n, {\rm i}\epsilon_{n'}; {\rm i}\nu_m)
	= \delta_{nn'} \chi^{{\rm c0}, \gamma}_{\mib{q}} ({\rm i}\epsilon_n; {\rm i}\nu_m) \nonumber \\
	&+ T\sum_{n''} \delta_{nn'} \chi^{{\rm c0}, \gamma}_{\mib{q}} ({\rm i}\epsilon_n; {\rm i}\nu_m)
	\Gamma^{{\rm c}, \gamma} ({\rm i}\epsilon_n, {\rm i}\epsilon_{n''}; {\rm i}\nu_m)
	 \chi^{{\rm c}, \gamma}_{\mib{q}} ({\rm i}\epsilon_{n''}, {\rm i}\epsilon_{n'}; {\rm i}\nu_m),
\label{eq:chi_c_q}
\end{align}
where $\chi^{{\rm c0}, \gamma}_{\mib{q}}$ is defined by $\chi^{{\rm c0}, \gamma}_{\mib{q}}=N C_N^{\gamma} \Pi_{\mib{q}}$ with
\begin{align}
	\Pi_{\mib{q}} ({\rm i}\epsilon_n; {\rm i}\nu_m)
	= -\frac{1}{N_0} \sum_{\mib{k}} G_{\mib{k}}^{\rm c} ({\rm i}\epsilon_n)
	 G_{\mib{k}+\mib{q}}^{\rm c} ({\rm i}\epsilon_n+{\rm i}\nu_m).
\label{eq:def_Pi_q}
\end{align}
Eliminating $\Gamma^{{\rm c}, \gamma}$ from eqs.~(\ref{eq:chi_c_loc}) and (\ref{eq:chi_c_q}), we obtain the following matrix equation for $\chi_{\mib{q}}^{\rm c}$:
\begin{align}
	[\chi^{{\rm c}, \gamma}_{\mib{q}}]^{-1} = [\chi^{{\rm c}, \gamma}_{\rm loc}]^{-1} - [\chi^{{\rm c0}, \gamma}_{\rm loc}]^{-1} + [\chi^{{\rm c0}, \gamma}_{\mib{q}}]^{-1}. 
\label{eq:chi_c_matrix}
\end{align}
This equation is identical to that in the Hubbard model\cite{Jarrell-Hubbard}.
This follows from the fact that all the spatial dependence can be represented by $G^{\rm c}_{\mib{k}}$, and therefore by $\Pi_{\mib{q}}$, in both the Anderson lattice and the CS lattice. 
Equations~(\ref{eq:chi_c_loc}) and (\ref{eq:chi_c_q}) represent effective onsite interactions between conduction electrons resulting from the interaction with the local moment. 
The total susceptibility including $\langle c^{\dag}_{\alpha} c_{\alpha} f^{\dag}_{\alpha'} f_{\alpha'} \rangle$ can also be derived in an analogous way\cite{Jarrell-AL}.

We should note that the DMFT guarantees self-consistency only in the single-particle Green function. 
Consequently the two-particle correlation function does not satisfy self-consistency between the effective impurity and the lattice models. 
Namely, eq.~(\ref{eq:chi_c_matrix}) gives
\begin{align}
	\chi_{\rm loc}^{\rm c} \neq \frac{1}{N_0} \sum_{\mib{q}} \chi_{\mib{q}}^{\rm c},
\end{align}
as in the ordinary random phase approximation (RPA) theory.
Hence, the present formalism corresponds to the RPA level in momentum  
dependence,
but proper local dynamics is incorporated through the irreducible  
vertex.

\section{Evaluation of the Generalized $t$-matrix in the CT-QMC}
Recently, a new impurity solver called CT-QMC has been developed\cite{Rubtsov, Werner, Otsuki-CTQMC}.
The single- and two-particle correlations in the impurity CS model can be obtained accurately within statistical errors\cite{Otsuki-CTQMC}.
In this section, we describe how to evaluate the generalized $t$-matrix of the impurity CS model in the CT-QMC.
It will be shown that the high-frequency limit for Green functions can be taken strictly in this formula.


To establish notations, we begin by summarizing the formula of the single-particle Green function. 
In the Monte Carlo simulations, a snap shot is expressed by an integer $k$ and two sets of variables, 
$(\tau_1, \cdots, \tau_k)$ for time and $(\alpha_1, \cdots, \alpha_k)$ for spin, 
which we symbolically represent by $\mathcal{K}$ hereafter.
For a given configuration of $\mathcal{K}$, the Green function is given by
\begin{align}
	G_{\alpha}(\tau, \tau'; \mathcal{K})
	&= g_{\alpha}(\tau-\tau') \nonumber \\
	&- \sum_{ij} g_{\alpha}(\tau-\tau_j) (M_{\alpha})_{ji} g_{\alpha}(\tau_i-\tau'),
\end{align}
where the matrix $M_{\alpha}$ is defined by $M_{\alpha} =D_{\alpha}^{-1}$
and 
$(D_{\alpha})_{ij}=g(\tau_i'' - \tau_j')$ with $\{ \tau_i'' \}$ and $\{ \tau_j' \}$ being certain sets of $\tau$'s in $\mathcal{K}$.
The average over a Monte Carlo ensemble gives the physical Green function: $G(\tau, \tau')=\langle G(\tau, \tau'; \mathcal{K}) \rangle_{\rm MC}$.

We consider the generalized susceptibility for conduction electrons defined by
\begin{align}
	\chi^{\rm c}_{\alpha \alpha'}(\tau_1, \tau_2&, \tau_3, \tau_4)
	= \langle T_{\tau} c_{\alpha}^{\dag}(\tau_1) c_{\alpha}(\tau_2) c_{\alpha'}^{\dag}(\tau_3) c_{\alpha'}(\tau_4) \rangle \nonumber \\
	&- \langle T_{\tau} c_{\alpha}^{\dag}(\tau_1) c_{\alpha}(\tau_2) \rangle
	  \langle T_{\tau} c_{\alpha'}^{\dag}(\tau_3) c_{\alpha'}(\tau_4) \rangle.
\end{align}
For each configuration of $\mathcal{K}$, the Wick's theorem is applicable to the four-operator average, so that we obtain 
\begin{align}
	\langle T_{\tau}& c_{\alpha}^{\dag}(\tau_1) c_{\alpha}(\tau_2)
	 c_{\alpha'}^{\dag}(\tau_3) c_{\alpha'}(\tau_4) ; \mathcal{K} \rangle \nonumber \\
	&= G_{\alpha}(\tau_2, \tau_1; \mathcal{K}) G_{\alpha'}(\tau_4, \tau_3; \mathcal{K}) \nonumber \\
	&- \delta_{\alpha \alpha'} G_{\alpha}(\tau_4, \tau_1; \mathcal{K}) G_{\alpha}(\tau_2, \tau_3; \mathcal{K}). 
\end{align}
This equation can actually be demonstrated with use of the fast-update formula for the four-operator addition\cite{Rubtsov}. 
Performing the Fourier transform defined by eq.~(\ref{eq:Fourier_chi_c_1234}) and taking the Monte Carlo average, we obtain a formula for $\chi^{\rm c}_{\alpha \alpha'}({\rm i}\epsilon_n, {\rm i}\epsilon_{n'}; {\rm i}\nu_m)$.
The generalized $t$-matrix $\mathcal{T}_{\alpha \alpha'} ({\rm i}\epsilon_n, {\rm i}\epsilon_{n'}; {\rm i}\nu_m)$, defined in eq.~(\ref{eq:chi_c_1234_gen_t}), is then given by
\begin{align}
	&\mathcal{T}_{\alpha \alpha'} ({\rm i}\epsilon_n, {\rm i}\epsilon_{n'}; {\rm i}\nu_m)
	= \big< u_{\alpha}(\epsilon_n, \epsilon_n + \nu_m) u_{\alpha'}(\epsilon_{n'}+\nu_m, \epsilon_{n'})
	\nonumber \\
	&\qquad - \delta_{\alpha \alpha'} u_{\alpha}(\epsilon_n, \epsilon_{n'}) u_{\alpha}(\epsilon_{n'}+\nu_m, \epsilon_n + \nu_m) \big>_{\rm MC} \nonumber \\
	&\quad - \delta_{m 0} t_{\alpha}({\rm i}\epsilon_n) t_{\alpha'}({\rm i}\epsilon_{n'}), 
\label{eq:MC_gen_T_omega}
\end{align}
where
\begin{align}
	u_{\alpha}(\epsilon_1, \epsilon_2)
	= T \sum_{ij} (M_{\alpha})_{ji} {\rm e}^{{\rm i}\epsilon_2 \tau_j - {\rm i}\epsilon_1 \tau_i}.
\end{align}
The impurity $t$-matrix is given in terms of $u_{\alpha}$ by $t_{\alpha}({\rm i}\epsilon_n) = -\langle u_{\alpha}(\epsilon_n, \epsilon_n) \rangle_{\rm MC}$.

The generalized $t$-matrix, given in the frequency domain, can also be expressed in the imaginary-time domain. 
From eq.~(\ref{eq:MC_gen_T_omega}), we obtain
\begin{align}
	&\mathcal{T}_{\alpha \alpha'}(\tau_1, \tau_2, \tau_3, \tau_4) \nonumber \\
	&= \Bigg< 
	\sum_{ijlm} [ (M_{\alpha})_{ji} (M_{\alpha'})_{ml}
	- \delta_{\alpha \alpha'} (M_{\alpha})_{jl} (M_{\alpha})_{mi} ] \nonumber \\
	&\qquad \times \delta(\tau_1-\tau_i) \delta(\tau_2-\tau_j) \delta(\tau_3-\tau_l) \delta(\tau_4-\tau_m)
	\Bigg>_{\rm MC} \nonumber \\
	&- t_{\alpha}(\tau_2 -\tau_1) t_{\alpha'}(\tau_4 -\tau_3).
\label{eq:MC_gen_T_tau}
\end{align}
Although it is in principle possible to measure $\mathcal{T}$ in the imaginary-time domain, it is not practical due to its complex structure. 
Namely, two discontinuities for each variable between 0 and $\beta$ make a precise treatment difficult. 
Instead, eq.~(\ref{eq:MC_gen_T_tau}) can be used for evaluating each term in eq.~(\ref{eq:def_gen_t}) selectively. 
For example, by letting $\tau_1=\tau_2$ and $\tau_3=\tau_4$ in eq.~(\ref{eq:MC_gen_T_tau}), the third term of eq.~(\ref{eq:def_gen_t}) is singled out to give
\begin{align}
	&\langle T_{\tau} f_{\alpha}^{\dag}(\tau) f_{\alpha}(\tau) f_{\alpha'}^{\dag} f_{\alpha'} \rangle \nonumber \\
	&= \frac{T}{J^2} \Bigg< 
	\sum_{\tau_i=\tau_j} \sum_{\tau_l=\tau_m} [ (M_{\alpha})_{ji} (M_{\alpha'})_{ml} \nonumber \\
	&\quad - \delta_{\alpha \alpha'} (M_{\alpha})_{jl} (M_{\alpha})_{mi} ] \delta_{+}(\tau, \tau_i-\tau_l) 
	\Bigg>_{\rm MC}, 
\label{eq:suscep_longitudinal}
\end{align}
where the $\beta$-periodicity is imposed on $\delta_+(\tau, \tau')$:
\begin{align}
	\delta_+(\tau, \tau') = 
	\left\{ \begin{array}{ll}
	\delta(\tau-\tau'), & (\tau'>0) \\
	\delta(\tau-\tau'-\beta). & (\tau'<0)
	\end{array} \right.
\end{align} 
The summations in eq.~(\ref{eq:suscep_longitudinal}) are taken for all pairs satisfying $\tau_i=\tau_j$ and $\tau_l=\tau_m$. 
On the other hand, by letting $\tau_1=\tau_4$ and $\tau_2=\tau_3$, eq.~(\ref{eq:MC_gen_T_tau}) yields the transverse susceptibility as
\begin{align}
	&\langle T_{\tau} f_{\alpha}^{\dag}(\tau) f_{\alpha'}(\tau) f_{\alpha'}^{\dag} f_{\alpha} \rangle \nonumber \\
	&= -\frac{T}{J^2} \Bigg< 
	\sum_{\tau_i=\tau_m} \sum_{\tau_j=\tau_l} [ (M_{\alpha})_{ji} (M_{\alpha'})_{ml} \nonumber \\
	&\quad - \delta_{\alpha \alpha'} (M_{\alpha})_{jl} (M_{\alpha})_{mi} ] \delta_{+}(\tau, \tau_i-\tau_l) 
	\Bigg>_{\rm MC}.
\label{eq:suscep_transverse}
\end{align}
As presented in ref.~\citen{Otsuki-CTQMC}, the susceptibility can also be evaluated from the configurations, $\{ \tau_i \}$ and $\{ \alpha_i \}$, without the matrix $M_{\alpha}$.
While eq.~(\ref{eq:suscep_longitudinal}) gives the same results as the configuration measurement within error bars, eq.~(\ref{eq:suscep_longitudinal}) tends to have more errors.
As for the transverse susceptibility, eq.~(\ref{eq:suscep_transverse}) is the only way of evaluation.

In order to obtain the spin correlations in the DMFT, we need the high-frequency limits of $\mathcal{T}_{\alpha \alpha'} ({\rm i}\epsilon_n, {\rm i}\epsilon_{n'}; {\rm i}\nu_m)$ as shown in \S\ref{sec:dmft}.
The high-frequency limit can be taken strictly in the same way as eqs.~(\ref{eq:suscep_longitudinal}) and (\ref{eq:suscep_transverse}).
In the limit $\epsilon_n \rightarrow \infty$, only terms with $\delta(\tau_1 - \tau_2)$ remain. 
Hence, this limit corresponds to restricting the summation in eq.~(\ref{eq:MC_gen_T_tau}) to $\tau_i=\tau_j$. 
After the Fourier transform, we obtain $\lim_{\epsilon_n \rightarrow \infty} \mathcal{T}_{\alpha \alpha'} ({\rm i}\epsilon_n, {\rm i}\epsilon_{n'}; {\rm i}\nu_m)$.
In a similar manner, the limit $\epsilon_{n'} \rightarrow \infty$ is calculated by restricting the summation to $\tau_l=\tau_m$.

\section{Structure of the Generalized $t$-matrix}

In this section, we show a typical example leading to the $\mib{q}$-dependent static susceptibilities.
We employ the Kondo lattice model in the form of eq.~(\ref{eq:H_KL2}).
Namely, the bare Green function includes the potential scattering $v$ as
$g_{\mib{k}}({\rm i}\epsilon_n)=({\rm i}\epsilon_n - \xi_{\mib{k}} - v)$.
We note that, in this definition, the $t$-matrix does not have the particle-hole symmetry even at the half filling.
We adopt the nearest-neighbor tight-binding band of the infinite-dimensional hyper-cubic lattice\cite{Muller-Hartmann}.
The density of states is given by $\rho(\omega)=D^{-1}\sqrt{2/\pi}\exp(-2\omega^2/D^2)$, and we take $D=1$ as the unit of energy.

\begin{figure}[tbp]
	\begin{center}
	\includegraphics[width=7cm]{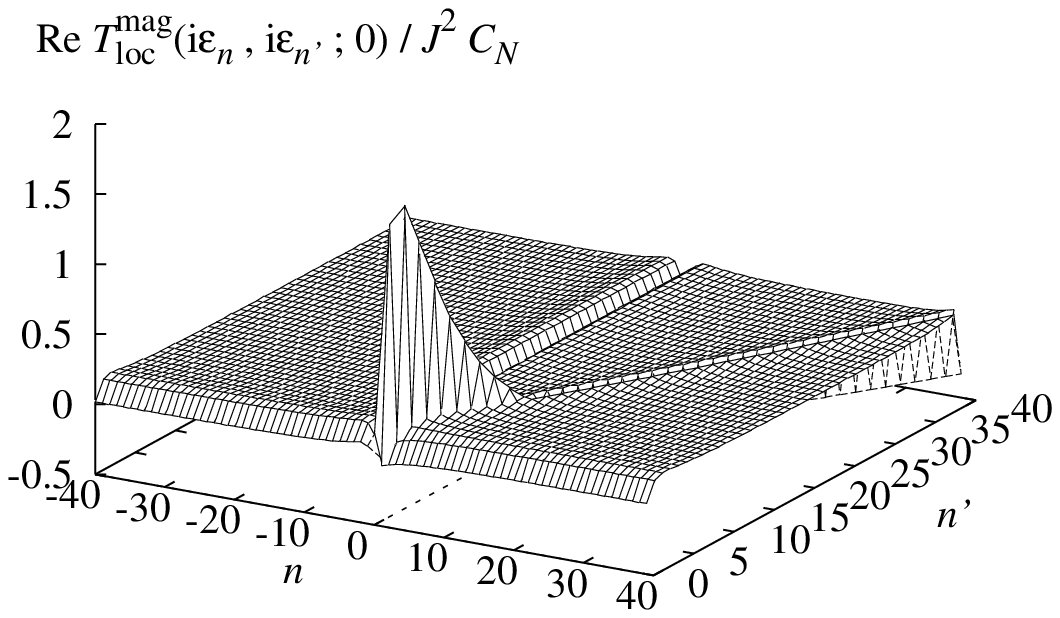}
	\includegraphics[width=7cm]{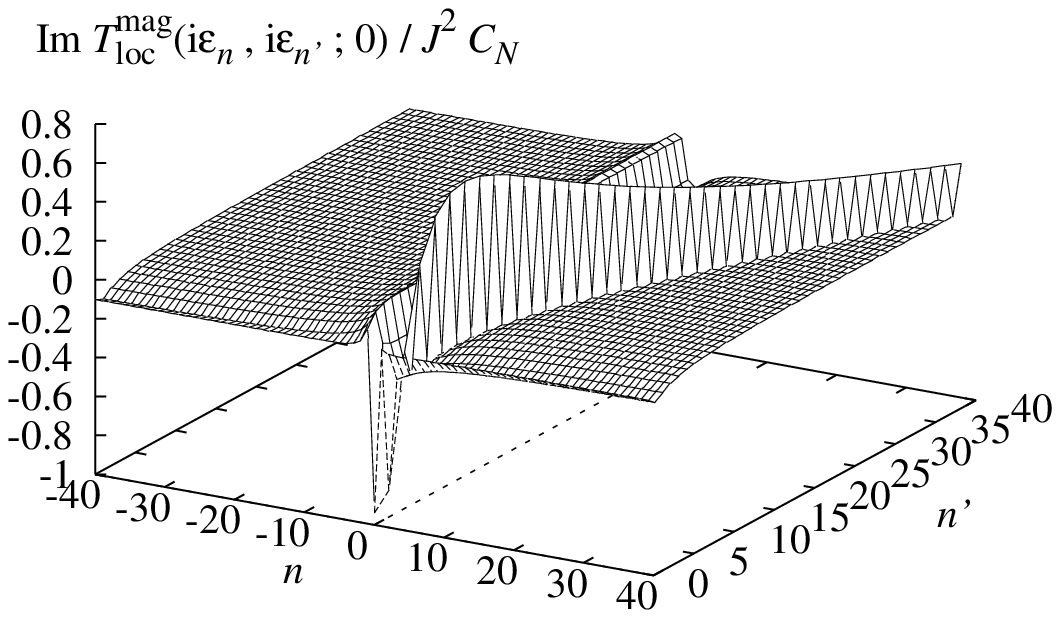}
	\end{center}
	\caption{$\mathcal{T}^{\rm mag}_{\rm loc}({\rm i}\epsilon_n, {\rm i}\epsilon_{n'}; 0)$ for $J=0.3$ and $T=0.02$.}
	\label{fig:KL-T_loc_mag}
\end{figure}

\begin{figure}[tbp]
	\begin{center}
	\includegraphics[width=8cm]{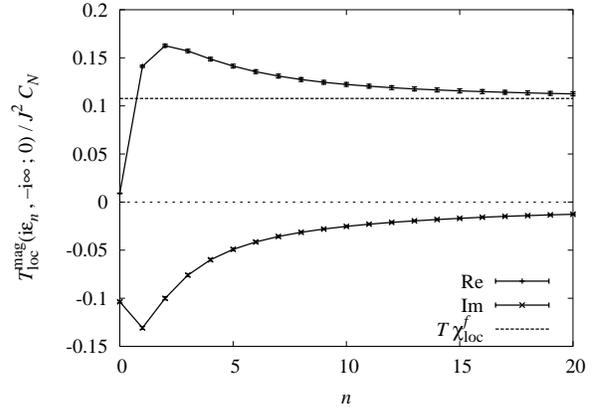}
	\end{center}
	\caption{$\mathcal{T}_{\rm loc}({\rm i}\epsilon_n, -{\rm i}\infty; 0)$ for $J=0.3$ and $T=0.02$.}
	\label{fig:KL-T_loc_inf}
\end{figure}

In the effective impurity problem, we evaluate the static component, ${\rm i}\nu_m=0$, of the generalized $t$-matrix of each channel $\gamma$,  $\mathcal{T}_{\rm loc}^{\gamma}({\rm i}\epsilon_n, {\rm i}\epsilon_{n'}; 0)$. 
Figure~\ref{fig:KL-T_loc_mag} shows the magnetic channel $\mathcal{T}_{\rm loc}^{\rm mag}({\rm i}\epsilon_n, {\rm i}\epsilon_{n'}; 0)$ for $J=0.3$ and $T=0.02$ at half filling. 
\begin{figure}[tbp]
	\begin{center}
	\includegraphics[width=7cm]{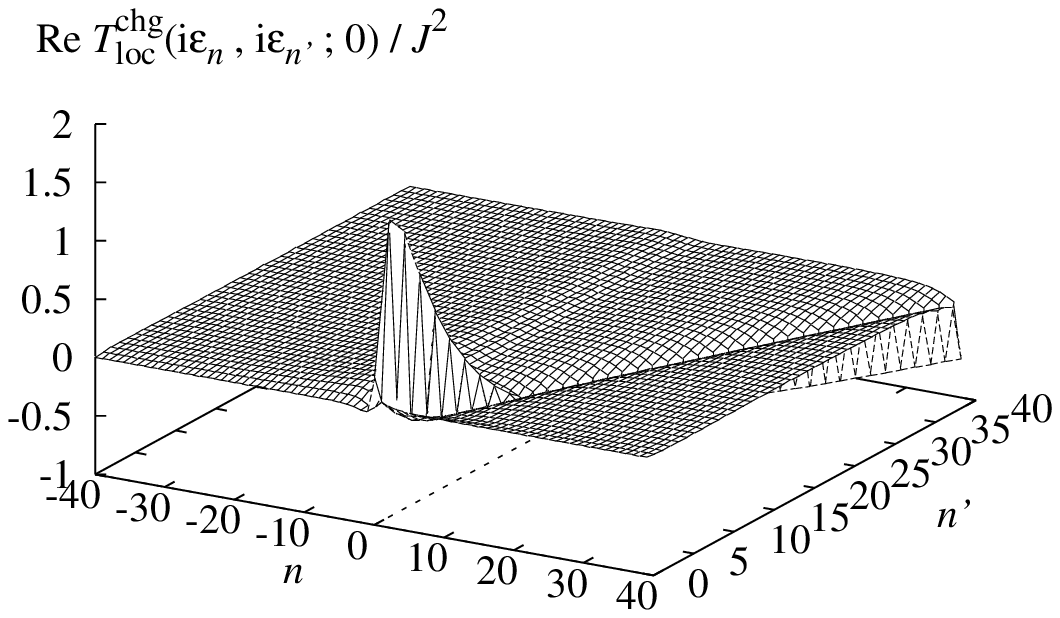}
	\includegraphics[width=7cm]{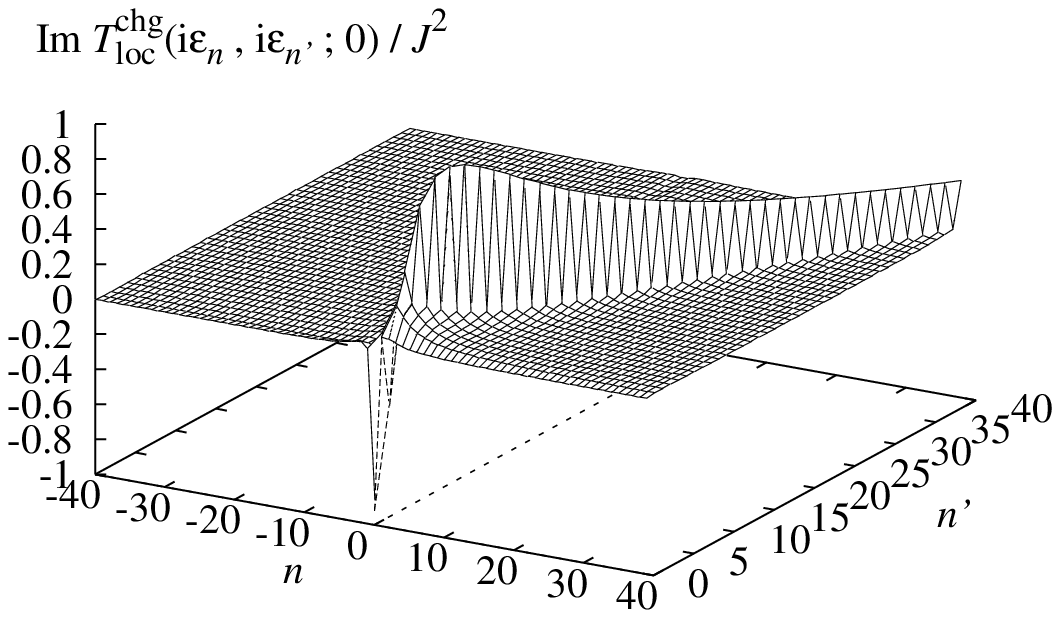}
	\end{center}
	\caption{$\mathcal{T}^{\rm chg}_{\rm loc}({\rm i}\epsilon_n, {\rm i}\epsilon_{n'}; 0)$ for $J=0.3$ and $T=0.02$.}
	\label{fig:KL-T_loc_chg}
\end{figure}
We plot only the range $n' \geq 0$ because of the relation $\mathcal{T}_{\rm loc}^{\gamma}(-{\rm i}\epsilon_n, -{\rm i}\epsilon_{n'}; 0) = \mathcal{T}_{\rm loc}^{\gamma}({\rm i}\epsilon_n, {\rm i}\epsilon_{n'}; 0)^*$.
The frequencies are taken up to $n=40$ to see the whole structure of the function. 
As shown later, much fewer elements are actually sufficient for the evaluation of $\chi^f_{\mib{q}}$.
We notice, in Fig.~\ref{fig:KL-T_loc_mag}, distinct structures around the diagonal, $n=n'$, and the zero frequencies, $n=0$ or $n'=0$. 
They are related to the $\delta$-functions in eq.~(\ref{eq:def_gen_t}), which do not disappear even in the high-frequency limit. 
The frequencies can be exactly taken to infinity in the CT-QMC. 
Figure~\ref{fig:KL-T_loc_inf} shows $\mathcal{T}_{\rm loc}({\rm i}\epsilon_n, -{\rm i}\infty; 0)$ for the same parameters as above.
While the function has some structure around $n=0$, at high frequencies we confirm a convergence to $T \chi^f_{\rm loc}$, which is provided in eq.~(\ref{eq:chi_f_1234_CS}). 
On the other hand, the charge channel $\mathcal{T}_{\rm loc}^{\rm chg}({\rm i}\epsilon_n, {\rm i}\epsilon_{n'}; 0)$ converges to zero at high frequencies, except for around the diagonal, $n=n'$.
This high-frequency behavior follows because there is no charge degree of freedom in the localized electrons in the Kondo lattice model.

\begin{figure}[tbp]
	\begin{center}
	\includegraphics[width=7cm]{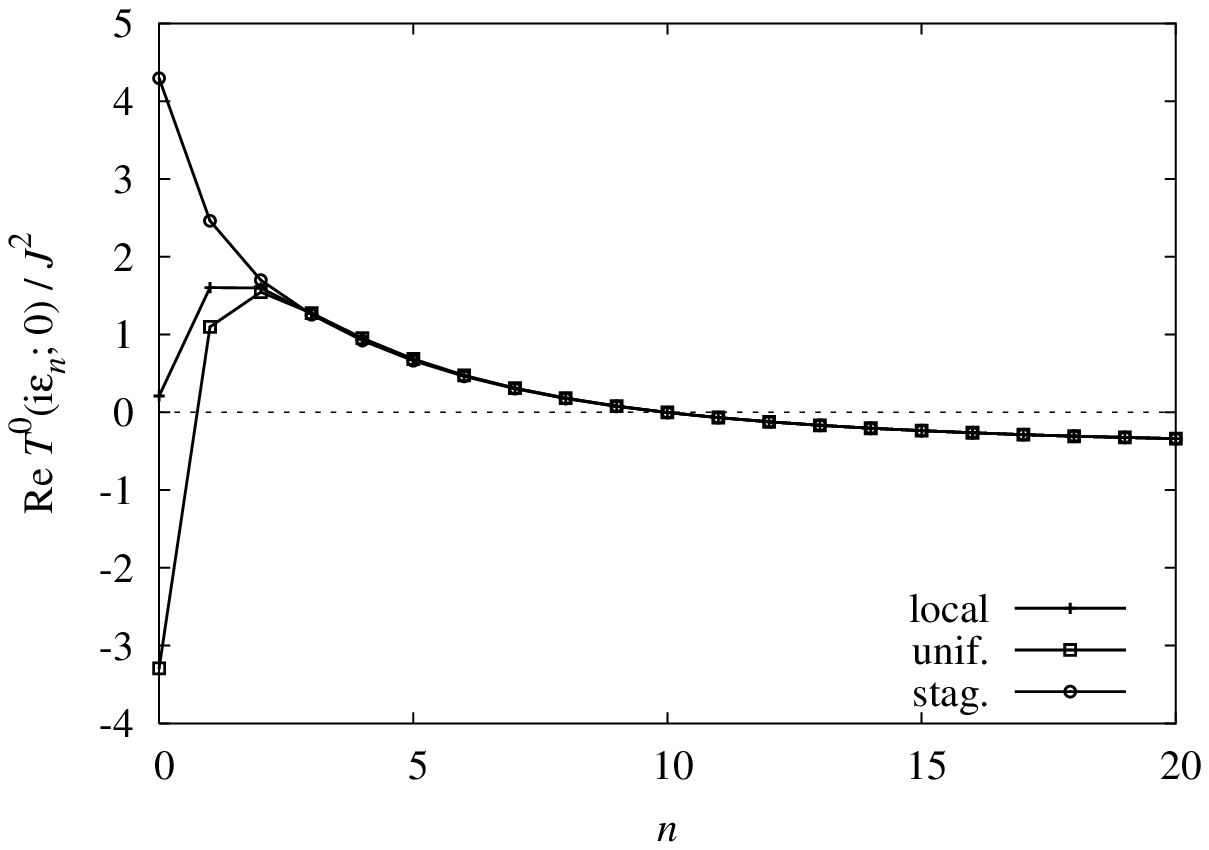}
	\includegraphics[width=7cm]{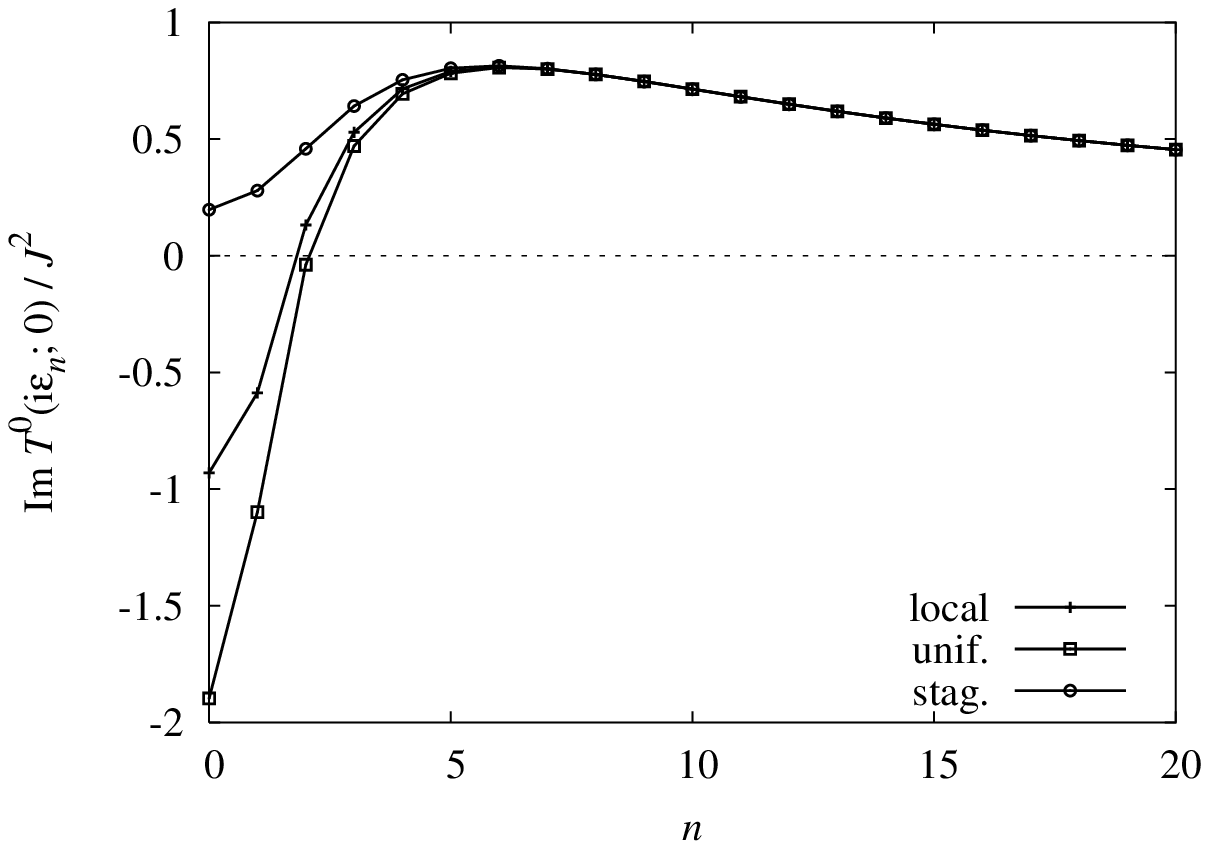}
	\end{center}
	\caption{Local, uniform ($\mib{q}=0$) and staggered ($\mib{q}=\mib{Q}$) components of $\mathcal{T}^0_{\mib{q}}({\rm i}\epsilon_n; 0)$ for $J=0.3$ and $T=0.02$. }
	\label{fig:KL-T0}
\end{figure}

\begin{figure}[tbp]
	\begin{center}
	\includegraphics[width=7cm]{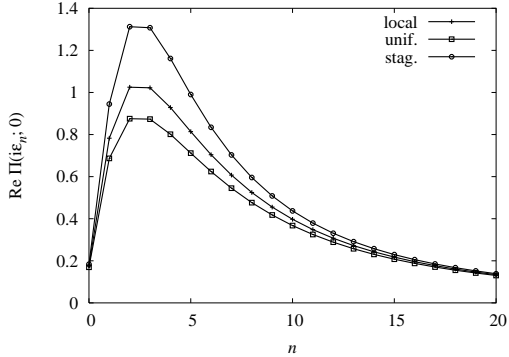}
	\end{center}
	\caption{Local, uniform and staggered components of $\Pi_{\mib{q}}({\rm i}\epsilon_n; 0)$ for $J=0.3$ and $T=0.02$. }
	\label{fig:KL-Pi}
\end{figure}

In order to compute $\mib{q}$-dependences of the generalized $t$-matrix with eq.~(\ref{eq:T_q_matrix}), we need the lowest order quantities without the local vertex. 
Figure~\ref{fig:KL-T0} shows $\mathcal{T}^0_{\rm loc}({\rm i}\epsilon_n; 0)$, $\mathcal{T}^0_{\mib{q}=0}({\rm i}\epsilon_n; 0)$ and $\mathcal{T}^0_{\mib{Q}}({\rm i}\epsilon_n; 0)$ defined in eqs.~(\ref{eq:T0_loc}) and (\ref{eq:T0_q}), 
where $\mib{Q}=(\pi, \cdots, \pi)$ representing the corner of the hyper-cubic Brillouin zone. 
The evaluation involves the $t$-matrices given in eqs.~(\ref{eq:t_site-diag}) and (\ref{eq:t_k}), and $\Pi_{\mib{q}}({\rm i}\epsilon_n; 0)$ defined in eq.~(\ref{eq:def_Pi_q}). 
The imaginary part of $\Pi_{\mib{q}}({\rm i}\epsilon_n; 0)$ is zero due to the particle-hole symmetry at $\nu_m=0$, and the real part is shown in Fig.~\ref{fig:KL-Pi}. 
From Fig.~\ref{fig:KL-T0}, the lowest order functions turn out to be almost identical with each other except for low frequencies. 
Hence, the difference of their inverses, $P_{\mib{q}} = \mathcal{T}^0_{\rm loc}({\rm i}\epsilon_n; 0)^{-1} - \mathcal{T}^0_{\mib{q}}({\rm i}\epsilon_n; 0)^{-1}$, vanishes rapidly with increasing $n$. 
This feature of $P_{\mib{q}}$ ensures an efficient and precise evaluation of the infinite-size matrix equation, eq.~(\ref{eq:T_q_matrix}). 
Only the low-frequency part where $P_{\mib{q}}$ has finite values is to be provided to obtain the dynamical susceptibilities of the local moments (see Appendix\ref{sec:high_freq_limit} for detail). 
This idea is also applicable to evaluations of the conduction-electron susceptibility from $\Pi_{\mib{q}}({\rm i}\epsilon_n; 0)$.

\section{Summary}

We have developed a framework to deal with the dynamics of highly correlated lattice models with localized electrons, such as the Kondo and the CS lattice models. 
The $t$-matrix, which describes the effect of the localized moments on the single-particle excitation, has been extended to the two-particle responses. 
The ordinary and the generalized $t$-matrices play essential roles in applying the conventional diagrammatical approach to models in the atomic limit. 
Consequently, the spatially dependent susceptibilities of the localized moments have been given in terms of the generalized $t$-matrix based on the DMFT. 

Our approach deals with the localized degrees of freedom from the strong correlation limit.  
The CT-QMC is very powerful as an impurity solver which gives finite-temperature dynamics of the Kondo-type model.
A numerical work in this framework will be presented in another publication.

\section*{Acknowledgment}
One of the authors (J.O.) was supported by Research Fellowships of the Japan Society for the Promotion of Science for Young Scientists.

\appendix

\section{Derivation of Expression of the Generalized $t$-matrix}
\label{sec:deriv_gen_t}

In this Appendix, we derive equations of motion of the single-particle and two-particle Green function, 
and demonstrate the expressions of the $t$-matrix and the generalized $t$-matrix.

\subsection{Some useful relations}
\label{useful-relations}
We begin with summarizing some relations 
which are necessary in taking derivative involving the time-ordering operator. 
Letting $A$ to $D$ be the fermion or boson, the two- and four-operator products are differentiated as
\begin{align}
	&\frac{\partial}{\partial \tau_1} \langle T_{\tau} A(\tau_1) B(\tau_2) \rangle
	= \left< T_{\tau} \frac{\partial A(\tau_1)}{\partial \tau_1} B(\tau_2) \right> \nonumber \\
	&\qquad + \delta(\tau_1 - \tau_2) \langle T_{\tau} [A(\tau_1), B(\tau_1) ]_{\pm} \rangle,
\label{eq:deriv_AB}
\end{align}
and
\begin{align}
	&\frac{\partial}{\partial \tau_1} \langle T_{\tau} A(\tau_1) B(\tau_2) C(\tau_3) D(\tau_4) \rangle \nonumber \\
	&= \left< T_{\tau} \frac{\partial A(\tau_1)}{\partial \tau_1} B(\tau_2) C(\tau_3) D(\tau_4) \right> \nonumber \\
	&\quad + \delta(\tau_1 -\tau_2) \langle T_{\tau} [ A(\tau_1), B(\tau_1) ]_{\pm} C(\tau_3) D(\tau_4) \rangle \nonumber \\
	&\quad + \delta(\tau_1 -\tau_3) \langle T_{\tau} [ A(\tau_1), C(\tau_1) ]_{\pm} D(\tau_4) B(\tau_2) \rangle \nonumber \\
	&\quad + \delta(\tau_1 -\tau_4) \langle T_{\tau} [ A(\tau_1), D(\tau_1) ]_{\pm} B(\tau_2) C(\tau_3) \rangle,
\label{eq:deriv_ABCD}
\end{align}
where $[A,B]_{\pm}=AB \pm BA$ is an anti-commutator and a commutator, and are chosen for fermion and boson, respectively. 
The following relation is satisfied when $A$ and $B$ are fermion operators and $\phi$ a boson operator:
\begin{align}
	&\frac{\partial}{\partial \tau_1} \langle T_{\tau} A(\tau_1) B(\tau_2) \phi(\tau_3) \rangle \nonumber \\
	& = \left< T_{\tau} \frac{\partial A(\tau_1)}{\partial \tau_1} B(\tau_2) \phi(\tau_3) \right> \nonumber \\
	&\quad + \delta(\tau_1 - \tau_2) \langle T_{\tau} \{A(\tau_1), B(\tau_1) \} \phi(\tau_3) \rangle \nonumber \\
	&\quad + \delta(\tau_1 - \tau_3) \langle T_{\tau} [A(\tau_1), \phi(\tau_1) ] B(\tau_2) \rangle.
\end{align}

%

\subsection{Single-particle Green function}
The single-particle Green function is defined by
\begin{align}
	G_{12}(\tau_1, \tau_2) = - \langle T_{\tau} c_1 (\tau_1) c_2^{\dag} (\tau_2) \rangle.
\end{align}
We first differentiate with respect to $\tau_1$. 
Using eq.~(\ref{eq:deriv_AB}) and substituting eq.~(\ref{eq:motion_c}), we obtain
\begin{align}
	&\left( -\frac{\partial}{\partial \tau_1} -\xi_1 \right) G_{12}(\tau_1, \tau_2) \nonumber \\
	&\qquad = \delta_{12} \delta(\tau_1-\tau_2) - \langle T_{\tau} j_1 (\tau_1) c_2^{\dag}(\tau_2) \rangle.
\end{align}
The second term is differentiated with respect to $\tau_2$ to yield
\begin{align}
	&\left( \frac{\partial}{\partial \tau_2} -\xi_2 \right) \langle T_{\tau} j_1 (\tau_1) c_2^{\dag}(\tau_2) \rangle \nonumber \\
	&\qquad = \langle T_{\tau} j_1 (\tau_1) j_2^{\dag}(\tau_2) \rangle
	- \delta(\tau_1-\tau_2) \langle \{ j_1, c_2^{\dag} \} \rangle.
\end{align}
After the Fourier transform, defined by eq.~(\ref{eq:Fourier_G}), we obtain
\begin{align}
	G_{12}({\rm i}\epsilon_n) = \delta_{12} g_1({\rm i}\epsilon_n)
	+ g_1({\rm i}\epsilon_n) t_{12}({\rm i}\epsilon_n) g_2({\rm i}\epsilon_n),
\label{eq:G_12_t_12}
\end{align}
where the $t$-matrix is defined by
\begin{align}
	t_{12}(\tau_1, \tau_2) = -\langle T_{\tau} j_1 (\tau_1) j_2^{\dag}(\tau_2) \rangle
	+ \delta(\tau_1-\tau_2) \langle \{ j_1 , c_2^{\dag} \} \rangle.
\end{align}
This equation yields eq.~(\ref{eq:t_matrix_imp}) for the impurity model, and eq.~(\ref{eq:t_matrix_perio}) for the periodic model.

\subsection{Two-particle Green function}
We proceed to the two-particle Green function in eq.~(\ref{eq:def_chi_c_1234}).
As shown in the above derivation, 
time-derivative of $c(\tau)$ or $c^{\dag}(\tau)$ brings 
about $j(\tau)$ or $j^{\dag}(\tau)$ with extra terms. 
In the following, we in turn differentiate the conduction-electron operators until all the operators are 
replaced by $j$ and $j^{\dag}$. 

We first differentiate with respect to $\tau_1$. 
Noting the relation in eq.~(\ref{eq:deriv_ABCD}), the equation of motion is given by
\begin{align}
	&\left( \frac{\partial}{\partial \tau_1} -\xi_1 \right) 
	\langle T_{\tau} c_1^{\dag}(\tau_1) c_2(\tau_2) c_3^{\dag}(\tau_3) c_4(\tau_4) \rangle \nonumber \\
	&= \langle T_{\tau} j_1^{\dag}(\tau_1) c_2(\tau_2) c_3^{\dag}(\tau_3) c_4(\tau_4) \rangle \nonumber \\
	&+ \delta(\tau_1-\tau_2) \delta_{12} G_{43} (\tau_4, \tau_3)
	- \delta(\tau_1-\tau_4) \delta_{14} G_{23} (\tau_2, \tau_3).
\label{eq:cccc_tau1}
\end{align}
We introduce an operator $\mathcal{F}$ which gives the Fourier transform as
\begin{align}
	\mathcal{F} &\langle T_{\tau} c_1^{\dag}(\tau_1) c_2(\tau_2) c_3^{\dag}(\tau_3) c_4(\tau_4) \rangle \nonumber \\
	&= \frac{1}{\beta^2} \int_0^{\beta} {\rm d}\tau_1 \cdots \int_0^{\beta} {\rm d}\tau_4
	\langle T_{\tau} c_1^{\dag}(\tau_1) c_2(\tau_2) c_3^{\dag}(\tau_3) c_4(\tau_4) \rangle \nonumber \\
	&\qquad \times {\rm e}^{{\rm i}\epsilon_n (\tau_2-\tau_1)}
	{\rm e}^{{\rm i}\epsilon_{n'} (\tau_4-\tau_3)}
	{\rm e}^{{\rm i}\nu_m (\tau_2-\tau_3)}.
\end{align}
This transformation replaces the differential operators as follows:
\begin{align}
	\frac{\partial}{\partial \tau_1} \rightarrow {\rm i}\epsilon_n, \quad
	-\frac{\partial}{\partial \tau_2} \rightarrow {\rm i}\epsilon_n + {\rm i}\nu_m, \nonumber \\
	\frac{\partial}{\partial \tau_3} \rightarrow {\rm i}\epsilon_{n'} + {\rm i}\nu_m, \quad
	-\frac{\partial}{\partial \tau_4} \rightarrow {\rm i}\epsilon_{n'}.
\end{align}
Equation~(\ref{eq:cccc_tau1}) is then transformed to give
\begin{align}
	\mathcal{F} &\langle T_{\tau} c_1^{\dag}(\tau_1) c_2(\tau_2) c_3^{\dag}(\tau_3) c_4(\tau_4) \rangle \nonumber \\
	&= g_1({\rm i}\epsilon_n) [\mathcal{F} \langle T_{\tau} j_1^{\dag}(\tau_1) c_2(\tau_2) c_3^{\dag}(\tau_3) c_4(\tau_4) \rangle \nonumber \\
	&\quad + \delta_{m0} \delta_{12} G_{43} ({\rm i}\epsilon_{n'})
	- \delta_{nn'} \delta_{14} G_{23} ({\rm i}\epsilon_n + {\rm i}\nu_m)].
\label{eq:Fcccc}
\end{align}
We proceed to make further derivative with respect to other time variables 
$\tau_2$, $\tau_3$ and $\tau_4$.
We obtain the following equations:
\begin{align}
	\mathcal{F} &\langle T_{\tau} j_1^{\dag}(\tau_1) c_2(\tau_2) c_3^{\dag}(\tau_3) c_4(\tau_4) \rangle \nonumber \\
	&= g_2({\rm i}\epsilon_n +{\rm i}\nu_m) [\mathcal{F} \langle T_{\tau} j_1^{\dag}(\tau_1) j_2(\tau_2) c_3^{\dag}(\tau_3) c_4(\tau_4) \rangle \nonumber \\
	&\quad - \delta_{nn'} \delta_{23} g_4({\rm i}\epsilon_{n}) t_{41}({\rm i}\epsilon_n) \nonumber \\
	&\quad + \delta_{m0} \delta_{34} g_3({\rm i}\epsilon_{n'}+{\rm i}\nu_m) \langle \{ j_1^{\dag}, c_2 \} \rangle ] \nonumber \\
	&+ g_2({\rm i}\epsilon_n +{\rm i}\nu_m) g_3({\rm i}\epsilon_{n'}+{\rm i}\nu_m) g_4({\rm i}\epsilon_{n}) \nonumber \\
	&\quad \times \{ \mathcal{F} [ \delta(\tau_1-\tau_2) \langle T_{\tau} \{ j_1^{\dag}(\tau_1), c_2(\tau_1) \} j_3^{\dag}(\tau_3) j_4(\tau_4) \rangle ] \nonumber \\
	&\qquad + \mathcal{F} [ \delta(\tau_1-\tau_2)\delta(\tau_3-\tau_4)\langle \{ j_1^{\dag}(\tau_1), c_2(\tau_1) \} \{ j_3^{\dag}(\tau_3), c_4(\tau_3) \} \rangle ] \ \},
\label{eq:Fjccc}
\end{align}
\begin{align}
	\mathcal{F} &\langle T_{\tau} j_1^{\dag}(\tau_1) j_2(\tau_2) c_3^{\dag}(\tau_3) c_4(\tau_4) \rangle \nonumber \\
	&= g_3({\rm i}\epsilon_{n'}+{\rm i}\nu_m)
	\{ \mathcal{F} \langle T_{\tau} j_1^{\dag}(\tau_1) j_2(\tau_2) j_3^{\dag}(\tau_3) c_4(\tau_4) \rangle \nonumber \\
	&\quad + \mathcal{F}[ \delta(\tau_3-\tau_4) \delta_{34} \langle T_{\tau} j_1^{\dag}(\tau_1) j_2(\tau_2) \rangle ] \} \nonumber \\
	&- g_3({\rm i}\epsilon_{n'}+{\rm i}\nu_m) g_4({\rm i}\epsilon_{n'}) \nonumber \\
	&\quad \times \{ \mathcal{F}[ \delta(\tau_2-\tau_3) \langle T_{\tau} j_1^{\dag}(\tau_1) \{ j_2(\tau_3), c_3^{\dag}(\tau_3) \} j_4(\tau_4) \rangle ] \nonumber \\
	&\qquad + \mathcal{F}[ \delta(\tau_1-\tau_4) \delta(\tau_2-\tau_3) \langle T_{\tau} \{ j_1^{\dag}(\tau_1), c_4(\tau_1) \} \{ j_2(\tau_3), c_3^{\dag}(\tau_3) \} \rangle ] \ \},
\label{eq:Fjjcc}
\end{align}
%
%
\begin{align}
	\mathcal{F} &\langle T_{\tau} j_1^{\dag}(\tau_1) j_2(\tau_2) j_3^{\dag}(\tau_3) c_4(\tau_4) \rangle \nonumber \\
	&= g_4({\rm i}\epsilon_n)
	\{ \mathcal{F} \langle T_{\tau} j_1^{\dag}(\tau_1) j_2(\tau_2) j_3^{\dag}(\tau_3) j_4(\tau_4) \rangle \nonumber \\
	&\quad + \mathcal{F}[ \delta(\tau_1-\tau_4) \langle T_{\tau} \{ j_1^{\dag}(\tau_1), c_4(\tau_1) \} j_2(\tau_2) j_3^{\dag}(\tau_3) \rangle ] \nonumber \\
	&\quad + \mathcal{F}[ \delta(\tau_3-\tau_4) \langle T_{\tau} j_1^{\dag}(\tau_1) j_2(\tau_2) \{ j_3^{\dag}(\tau_3), c_4(\tau_3) \} \rangle ] \ \}.
\label{eq:Fjjjc}
\end{align}
%
%
%
Substituting eqs.~(\ref{eq:Fjccc})--(\ref{eq:Fjjjc}) into eq.~(\ref{eq:Fcccc}) and after some manipulations, 
we finally obtain an expression of $\mathcal{F} \langle T_{\tau} c_1^{\dag}(\tau_1) c_2(\tau_2) c_3^{\dag}(\tau_3) c_4(\tau_4) \rangle$ as follows:
\begin{align}
	\mathcal{F} &\langle T_{\tau} c_1^{\dag}(\tau_1) c_2(\tau_2) c_3^{\dag}(\tau_3) c_4(\tau_4) \rangle \nonumber \\
	&= \delta_{m0} [ \delta_{12} g_1({\rm i}\epsilon_n) \cdot G_{43} ({\rm i}\epsilon_{n'}) \nonumber \\
	 &\quad + g_2({\rm i}\epsilon_n) t_{21}({\rm i}\epsilon_n) g_1({\rm i}\epsilon_n) \cdot \delta_{34} g_4 ({\rm i}\epsilon_{n'}) ] \nonumber \\
	&- \delta_{nn'} [ \delta_{14} g_1({\rm i}\epsilon_n) \cdot G_{23} ({\rm i}\epsilon_{n} + {\rm i}\nu_m) \nonumber \\
	 &\quad + g_4({\rm i}\epsilon_n) t_{41}({\rm i}\epsilon_n) g_1({\rm i}\epsilon_n) \cdot \delta_{23} g_2 ({\rm i}\epsilon_{n} + {\rm i}\nu_m)] \nonumber \\
	&+ g_1({\rm i}\epsilon_n) g_2({\rm i}\epsilon_n + {\rm i}\nu_m) g_3({\rm i}\epsilon_{n'} + {\rm i}\nu_m) g_4({\rm i}\epsilon_{n'}) \nonumber \\
	&\quad \times \mathcal{F} \mathcal{T}'_{1234}(\tau_1, \tau_2, \tau_3, \tau_4),
\label{eq:Fcccc-1}
\end{align}
where $\mathcal{T}'_{1234}(\tau_1, \tau_2, \tau_3, \tau_4)$ is defined by
\begin{align}
	&\mathcal{T}'_{1234}(\tau_1, \tau_2, \tau_3, \tau_4)
	= \langle T_{\tau} j_1^{\dag}(\tau_1) j_2(\tau_2) j_3^{\dag}(\tau_3) j_4(\tau_4) \rangle \nonumber \\
	&+ \delta(\tau_1-\tau_2) \delta(\tau_3-\tau_4) \langle T_{\tau} \{ j_1^{\dag}(\tau_1), c_2(\tau_1) \}
	 \{ j_3^{\dag}(\tau_3), c_4(\tau_3) \} \rangle \nonumber \\
	&- \delta(\tau_1-\tau_4) \delta(\tau_2-\tau_3) \langle T_{\tau} \{ j_1^{\dag}(\tau_1), c_4(\tau_1) \}
	 \{ j_3^{\dag}(\tau_3), c_2(\tau_3) \} \rangle \nonumber \\
	&+ \delta(\tau_1-\tau_2) \langle T_{\tau} \{ j_1^{\dag}(\tau_1), c_2(\tau_1) \} j_3^{\dag}(\tau_3) j_4(\tau_4) \rangle \nonumber \\
	&+ \delta(\tau_3-\tau_4) \langle T_{\tau} j_1^{\dag}(\tau_1) j_2(\tau_2) \{ j_3^{\dag}(\tau_3), c_4(\tau_3) \} \rangle \nonumber \\
	&- \delta(\tau_1-\tau_4) \langle T_{\tau} \{ j_1^{\dag}(\tau_1), c_4(\tau_1) \} j_3^{\dag}(\tau_3) j_2(\tau_2) \rangle \nonumber \\
	&- \delta(\tau_2-\tau_3) \langle T_{\tau} j_1^{\dag}(\tau_1) j_4(\tau_4) \{ j_3^{\dag}(\tau_3), c_2(\tau_3) \} \rangle.
\end{align}
In this derivation, we have used $\{ j_2, c_3^{\dag} \} = \{ c_2, j_3^{\dag} \}$, which is demonstrated from the definition of $j_2$ and $j_3^{\dag}$. 
Noting eq.~(\ref{eq:G_12_t_12}), eq.~(\ref{eq:Fcccc-1}) is rewritten as
\begin{align}
	\mathcal{F} &\langle T_{\tau} c_1^{\dag}(\tau_1) c_2(\tau_2) c_3^{\dag}(\tau_3) c_4(\tau_4) \rangle
	- \delta_{m0} G_{21}({\rm i}\epsilon_n) G_{43} ({\rm i}\epsilon_{n'}) \nonumber \\
	&= - \delta_{nn'} [ \delta_{14} g_1({\rm i}\epsilon_n) \cdot G_{23} ({\rm i}\epsilon_{n} + {\rm i}\nu_m) \nonumber \\
	&\quad + g_4({\rm i}\epsilon_n) t_{41}({\rm i}\epsilon_n) g_1({\rm i}\epsilon_n) \cdot \delta_{23} g_2 ({\rm i}\epsilon_{n} + {\rm i}\nu_m)] \nonumber \\
	&+ g_1({\rm i}\epsilon_n) g_2({\rm i}\epsilon_n + {\rm i}\nu_m) g_3({\rm i}\epsilon_{n'} + {\rm i}\nu_m) g_4({\rm i}\epsilon_{n'}) \nonumber \\
	&\quad \times \mathcal{F} \mathcal{T}_{1234}(\tau_1, \tau_2, \tau_3, \tau_4),
\end{align}
where $\mathcal{T}_{1234}(\tau_1, \tau_2, \tau_3, \tau_4)$ is given by
\begin{align}
	&\mathcal{T}_{1234}(\tau_1, \tau_2, \tau_3, \tau_4) \nonumber \\
	&= \mathcal{T}'_{1234}(\tau_1, \tau_2, \tau_3, \tau_4) - t_{21}(\tau_2, \tau_1) t_{43}(\tau_4, \tau_3).
\end{align}
As a result, we obtain the expression of $\mathcal{T}_{1234}(\tau_1, \tau_2, \tau_3, \tau_4)$ in eq.~(\ref{eq:def_gen_t}).

\section{Internal Energy}
\label{sec:internal_energy}

In this appendix, we derive an expression of the internal energy for the impurity and periodic CS models. 
We begin by considering the CS lattice model in eq.~(\ref{eq:H_CSL}). 
Although the Hamiltonian include two-body interaction, its expectation value can be represented by the single-particle Green function with use of the technique introduced in ref.~\citen{Fetter-Walecka}. 
Considering an equation of motion for $G_{\mib{k}\alpha, \mib{k}\alpha} (\tau, \tau')$ with $\tau' \rightarrow \tau+0$, and then taking a summation for indices, we obtain
\begin{align}
	&-\lim_{\tau' \rightarrow \tau+0} \frac{\partial}{\partial \tau}
	\sum_{\mib{k}\alpha} G_{\mib{k}\alpha, \mib{k}\alpha} (\tau, \tau') \nonumber \\
	&= -\sum_{\mib{k}\alpha} \left< c^{\dag}_{\mib{k}\alpha} (\tau) \frac{\partial c_{\mib{k}\alpha} (\tau)}{\partial \tau} \right> \nonumber \\
	&= \left< \sum_{\mib{k}\alpha} \xi_{\mib{k}} c^{\dag}_{\mib{k}\alpha} c_{\mib{k}\alpha}
	+ J \sum_{i \alpha \alpha'} f^{\dag}_{i\alpha} f_{i\alpha'} c^{\dag}_{i\alpha'} c_{i\alpha}
	\right> \nonumber \\
	&= \langle H_{\rm CSL} \rangle - \mu \langle N_{\rm c} \rangle,
\end{align}
where we have used eq.~(\ref{eq:motion_c}). 
Taking the Fourier transform with respect to $\tau$, we obtain the internal energy $E=\langle H_{\rm CSL} \rangle$ as
\begin{align}
	E = T\sum_n \sum_{\mib{k}\alpha} {\rm i}\epsilon_n G_{\mib{k}\alpha, \mib{k}\alpha} ({\rm i}\epsilon_n) {\rm e}^{{\rm i}\epsilon_n \delta}
	 + \mu \langle N_{\rm c} \rangle,
\label{eq:energy_CSL}
\end{align}
where $\delta$ is a positive infinitesimal. 
In numerical calculations, we may subtract $1/{\rm i}\epsilon_n$ from the Green function, because it does not contribute to the internal energy. 
Furthermore, the quadratic term, $c/({\rm i}\epsilon_n)^2$, can be evaluated to be $c/2$.
By expanding eq.~(\ref{eq:G_c}) in powers of $1/z$, we obtain $c=J n^f_{\alpha} - \mu$ on condition of $\rho(\epsilon)=\rho(-\epsilon)$.
Hence, excepting this term, the series in eq.~(\ref{eq:energy_CSL}) converges as $({\rm i} \epsilon_n)^{-2}$ at high frequencies.

Equation~(\ref{eq:energy_CSL}) is also applicable to the impurity CS model. 
In this case, the total energy $E=\langle H_{\rm CS} \rangle$ is composed by the kinetic-energy part being proportional to $N_0$ and an impurity contribution. 
To distinguish the impurity contribution, we express $G_{\mib{k}\alpha, \mib{k}\alpha} ({\rm i}\epsilon_n)$ in terms of the impurity $t$-matrix $t_{\alpha}({\rm i}\epsilon_n)$ as follows:
\begin{align}
	G_{\mib{k}\alpha, \mib{k}\alpha} ({\rm i}\epsilon_n)
	= g_{\mib{k}} ({\rm i}\epsilon_n) +  g_{\mib{k}} ({\rm i}\epsilon_n) \frac{t_{\alpha} ({\rm i}\epsilon_n)}{N_0} g_{\mib{k}} ({\rm i}\epsilon_n).
\end{align}
Substituting this expression into eq.~(\ref{eq:energy_CSL}), we obtain the following equation for a change in the internal energy due to the impurity:
\begin{align}
	E_{\rm imp}&=
	\langle H_{\rm CS} \rangle - \langle H_{\rm c} \rangle_{\rm c} \nonumber \\
	&= T\sum_n {\rm i}\epsilon_n \left( \frac{1}{N_0} \sum_{\mib{k}\alpha} g^2_{\mib{k}\alpha} ({\rm i}\epsilon_n) \right)
	t_{\alpha}({\rm i}\epsilon_n) {\rm e}^{{\rm i}\epsilon_n \delta}.
\end{align}
Here, we have set $\mu=0$. 

\section{Implementation of High-Frequency Limit}
\label{sec:high_freq_limit}

As presented in \S\ref{sec:dmft}, the $\mib{q}$-dependent dynamical susceptibility $\chi^f_{\mib{q}} ({\rm i}\nu_m)$ in the CS lattice model has been given by the high-frequency limit of the generalized $t$-matrix $\mathcal{T}_{\mib{q}}({\rm i}\epsilon_n, {\rm i}\epsilon_{n'}; {\rm i}\nu_m)$.
To obtain $\mathcal{T}_{\mib{q}}$, we need to solve the infinite-size matrix equation, eq.~(\ref{eq:T_q_matrix}). 
In this appendix, we present a efficient way to deal with the infinite-size matrices accurately.

The matrix equation in eq.~(\ref{eq:T_q_matrix}) may be rewritten as
\begin{align}
	\mathcal{T}_{\mib{q}} = \mathcal{T}_{\rm loc}
	+ \mathcal{T}_{\rm loc} P_{\mib{q}} \mathcal{T}_{\rm loc}
	+ \mathcal{T}_{\rm loc} P_{\mib{q}} \mathcal{T}_{\rm loc} P_{\mib{q}} \mathcal{T}_{\rm loc} + \cdots,
\label{eq:T_q_appendix}
\end{align}
where $P_{\mib{q}}$ is a diagonal matrix whose element is defined by
\begin{align}
	(P_{\mib{q}})_l = \mathcal{T}_{\rm loc}^0 ({\rm i}\epsilon_l; {\rm i}\nu_m)^{-1}
	- \mathcal{T}_{\mib{q}}^0({\rm i}\epsilon_l; {\rm i}\nu_m)^{-1}.
\label{eq:def_P}
\end{align}
In eq.~(\ref{eq:def_P}), only a low-frequency part has finite values, since two terms cancel out each other at high frequencies.
Therefore, we can replace the infinite sum over $l$ with a summation up to $L$ 
\begin{align}
	\sum_{l=-\infty}^{+\infty} \rightarrow \sum_{|l| \leq L} \equiv {\sum_l}'.
\end{align}
In Fig.~\ref{fig:KL-T0}, for example, $L=10$ is sufficient for reliable calculations. 
With use of this restricted sum, eq.~(\ref{eq:T_q_appendix}) is rewritten as
\begin{align}
	(\mathcal{T}_{\mib{q}})_{nn'} = (\mathcal{T}_{\rm loc})_{nn'}
	+ {\sum_{ll'}}' (\mathcal{T}_{\rm loc})_{nl} (Q_{\mib{q}})_{ll'} (\mathcal{T}_{\rm loc})_{l'n'},
\end{align}
where the matrix $Q_{\mib{q}}$ is defined within the restricted space as follows:
\begin{align}
	Q_{\mib{q}} = P_{\mib{q}} + P_{\mib{q}} \mathcal{T}_{\rm loc} P_{\mib{q}} + \cdots
	= [1-P_{\mib{q}} \mathcal{T}_{\rm loc}]^{-1} P_{\mib{q}}.
\end{align}
Hence, in order to obtain $(\mathcal{T}_{\mib{q}})_{+\infty, -\infty}$, which yields $\chi_{\mib{q}}^f({\rm i}\nu_m)$ in eq.~(\ref{eq:T_q_limit}), we need to evaluate $(\mathcal{T}_{\rm loc})_{ll'}$ at low frequencies as well as
$(\mathcal{T}_{\rm loc})_{+\infty,l}$,
$(\mathcal{T}_{\rm loc})_{l',-\infty}$, and 
$(\mathcal{T}_{\rm loc})_{+\infty,-\infty}$.
The limit $\epsilon_n \rightarrow \infty$ in $\mathcal{T}_{\rm loc}$ can be taken strictly in the CT-QMC.
Consequently, the infinite-size matrix equation can be solved rigorously to give $\chi_{\mib{q}}^f({\rm i}\nu_m)$.

\section{Vertex Parts for Conduction Electrons}
\label{sec:vertex_cond}

We have introduced the generalized $t$-matrix $\mathcal{T}$ from the two-particle Green function of conduction electrons $\chi^{\rm c}$. 
With use of the integral equation for $\mathcal{T}$, the spatial dependence of $\mathcal{T}$ in the periodic system has been derived. 
In this appendix, we derive an integral equation for $\chi^{\rm c}$ itself. 

The source of the spatial dependence is only $\epsilon_{\mib{k}}$ in the Anderson lattice and the CS lattice models. 
Hence, we can represent all the spatial dependences explicitly by $G_{\mib{k}}^{\rm c}({\rm i}\epsilon_n)$, and therefore by $\Pi_{\mib{k} \mib{q}}({\rm i}\epsilon_n; {\rm i}\nu_m)=-G^{\rm c}_{\mib{k}}({\rm i}\epsilon_n) G^{\rm c}_{\mib{k}+\mib{q}}({\rm i}\epsilon_n +{\rm i}\nu_m)$. 
In the following, we omit the spin degeneracy for simplicity. 
The susceptibility of the conduction electrons $\chi^{\rm c}_{\mib{k}\mib{k}'\mib{q}}$ is given in terms of $\mathcal{T}_{\mib{k}\mib{k}'\mib{q}}$ in eq.~(\ref{eq:chi_c_gen_t}), 
which may be rewritten as
\begin{align}
	&\chi_{\mib{k} \mib{k}' \mib{q}}^{\rm c} ({\rm i}\epsilon_n, {\rm i}\epsilon_{n'}; {\rm i}\nu_m)
	= \delta_{nn'} \delta_{\mib{k}\mib{k}'} \Pi_{\mib{k} \mib{q}} ({\rm i}\epsilon_n; {\rm i}\nu_m) \nonumber \\
	&+ g_{\mib{k}}({\rm i}\epsilon_n) g_{\mib{k}+\mib{q}}({\rm i}\epsilon_n +{\rm i}\nu_m)
	 g_{\mib{k}'+\mib{q}}({\rm i}\epsilon_{n'} +{\rm i}\nu_m) g_{\mib{k}'}({\rm i}\epsilon_{n'}) \nonumber \\
	& \quad \times [ \mathcal{T}_{\mib{k} \mib{k}' \mib{q}} ({\rm i}\epsilon_n, {\rm i}\epsilon_{n'}; {\rm i}\nu_m)
	 - \delta_{nn'} \delta_{\mib{k}\mib{k}'} \mathcal{T}^{0}_{\mib{k} \mib{q}} ({\rm i}\epsilon_n; {\rm i}\nu_m)], 
\label{eq:chi_c_kkq}
\end{align}
where $\mathcal{T}^{0}_{\mib{k} \mib{q}}({\rm i}\epsilon_n; {\rm i}\nu_m)=-t_{\mib{k}}({\rm i}\epsilon_n) t_{\mib{k}+\mib{q}}({\rm i}\epsilon_n +{\rm i}\nu_m)$.
Using the irreducible vertex part $I({\rm i}\epsilon_n, {\rm i}\epsilon_{n'}; {\rm i}\nu_m)$ in eq.~(\ref{eq:T_q}), 
the terms in the bracket is expressed in the matrix form for $\epsilon_n$ and $\epsilon_{n'}$ as
\begin{align}
	\mathcal{T}_{\mib{k} \mib{k}' \mib{q}} - \delta_{\mib{k} \mib{k}'} \mathcal{T}^{0}_{\mib{k} \mib{q}}
	= \mathcal{T}^0_{\mib{k} \mib{q}} I'_{\mib{q}} \mathcal{T}^0_{\mib{k}' \mib{q}},
\label{eq:T_T0}
\end{align}
where $I'_{\mib{q}}({\rm i}\epsilon_n, {\rm i}\epsilon_{n'}; {\rm i}\nu_m)$ is a reducible vertex part composed of $I$ and $\mathcal{T}^0_{\mib{q}}$:
\begin{align}
	I'_{\mib{q}}
	&= I + I \mathcal{T}^0_{\mib{q}} I + I \mathcal{T}^0_{\mib{q}} I \mathcal{T}^0_{\mib{q}} I + \cdots.
\label{eq:I'_q}
\end{align}
From eqs.~(\ref{eq:Gc_k}) and (\ref{eq:t_k}), we can derive the relation $g_{\mib{k}}({\rm i}\epsilon_n) t_{\mib{k}}({\rm i}\epsilon_n) = G^{\rm c}_{\mib{k}}({\rm i}\epsilon_n) \Sigma^{\rm c}({\rm i}\epsilon_n)$, which leads to
\begin{align}
	&g_{\mib{k}}({\rm i}\epsilon_n) g_{\mib{k}+\mib{q}}({\rm i}\epsilon_n +{\rm i}\nu_m)
	\mathcal{T}^0_{\mib{k} \mib{q}} ({\rm i}\epsilon_n; {\rm i}\nu_m) \nonumber \\
	&\quad = \Pi_{\mib{k} \mib{q}} ({\rm i}\epsilon_n; {\rm i}\nu_m)
	\Sigma^{\rm c} ({\rm i}\epsilon_n) \Sigma^{\rm c} ({\rm i}\epsilon_n +{\rm i}\nu_m).
\label{eq:T0_Pi}
\end{align}
Substituting eqs.~(\ref{eq:T_T0}) and (\ref{eq:T0_Pi}) into eq.~(\ref{eq:chi_c_kkq}), and then taking the summation over $\mib{k}$ and $\mib{k}'$, 
we obtain 
\begin{align}
	&\chi_{\mib{q}}^{\rm c} ({\rm i}\epsilon_n, {\rm i}\epsilon_{n'}; {\rm i}\nu_m)
	= \Pi_{\mib{q}} ({\rm i}\epsilon_n; {\rm i}\nu_m) \delta_{nn'} \nonumber \\
	&\quad + \Pi_{\mib{q}} ({\rm i}\epsilon_n; {\rm i}\nu_m)
	 K'_{\mib{q}} ({\rm i}\epsilon_n, {\rm i}\epsilon_{n'}; {\rm i}\nu_m)
	 \Pi_{\mib{q}} ({\rm i}\epsilon_{n'}; {\rm i}\nu_m),
\label{eq:chi_c_q_K_q}
\end{align}
where
\begin{align}
	&K'_{\mib{q}}({\rm i}\epsilon_n, {\rm i}\epsilon_{n'}; {\rm i}\nu_m)
	= \Sigma^{\rm c}({\rm i}\epsilon_n) \Sigma^{\rm c}({\rm i}\epsilon_n + {\rm i}\nu_m) \nonumber \\
	&\qquad \times I'_{\mib{q}}({\rm i}\epsilon_n, {\rm i}\epsilon_{n'}; {\rm i}\nu_m)
	 \Sigma^{\rm c}({\rm i}\epsilon_{n'}) \Sigma^{\rm c}({\rm i}\epsilon_{n'} + {\rm i}\nu_m). 
\label{eq:def_K_q}
\end{align}

We focus on $I'_{\mib{q}}$ defined in eq.~(\ref{eq:I'_q}). 
The spatial dependence of $I'_{\mib{q}}$ originates in $\mathcal{T}^0_{\mib{q}}$, which can be expressed in terms of $\Pi_{\mib{q}}$. 
With use of the expression of $t_{\mib{k}}$ in eq.~(\ref{eq:t_k}), $\mathcal{T}^0_{\mib{q}}$ defined by eq.~(\ref{eq:T0_q}) is rewritten as
\begin{align}
	\mathcal{T}^0_{\mib{q}}({\rm i}\epsilon_n; {\rm i}\nu_m)
	&= \Sigma^{\rm c}({\rm i}\epsilon_n)^2
	[ A({\rm i}\epsilon_n; {\rm i}\nu_m) \nonumber \\
	&\qquad + \Pi_{\mib{q}}({\rm i}\epsilon_n; {\rm i}\nu_m) ]
	\Sigma^{\rm c}({\rm i}\epsilon_n + {\rm i}\nu_m)^2,
\end{align}
where $A({\rm i}\epsilon_n; {\rm i}\nu_m)$ is defined by
\begin{align}
	A({\rm i}\epsilon_n; {\rm i}\nu_m)
	&= \Sigma^{\rm c}({\rm i}\epsilon_n)^{-1} \Sigma^{\rm c}({\rm i}\epsilon_n + {\rm i}\nu_m)^{-1} \nonumber \\
	&+ \Sigma^{\rm c}({\rm i}\epsilon_n)^{-1} \bar{G}^{\rm c}({\rm i}\epsilon_n + {\rm i}\nu_m) \nonumber \\
	&+ \bar{G}^{\rm c}({\rm i}\epsilon_n) \Sigma^{\rm c}({\rm i}\epsilon_n + {\rm i}\nu_m)^{-1}.
\end{align}
It is clear from the above equation that $A$ composed only of local quantities. 
Introducing an auxiliary quantity $K$ by
\begin{align}
	&K ({\rm i}\epsilon_n, {\rm i}\epsilon_{n'}; {\rm i}\nu_m)
	= \Sigma^{\rm c}({\rm i}\epsilon_n) \Sigma^{\rm c}({\rm i}\epsilon_n + {\rm i}\nu_m) \nonumber \\
	&\qquad \times I ({\rm i}\epsilon_n, {\rm i}\epsilon_{n'}; {\rm i}\nu_m)
	 \Sigma^{\rm c}({\rm i}\epsilon_{n'}) \Sigma^{\rm c}({\rm i}\epsilon_{n'} + {\rm i}\nu_m), 
\label{eq:def_K}
\end{align}
$K'_{\mib{q}}$ in eq.~(\ref{eq:def_K_q}) is rewritten as
\begin{align}
	K'_{\mib{q}} &= K + K (A+ \Pi_{\mib{q}}) K'_{\mib{q}} \nonumber \\
	&= \Gamma^{\rm c} + \Gamma^{\rm c} \Pi_{\mib{q}} K'_{\mib{q}},
\label{eq:K_Gamma_c}
\end{align}
where $\Gamma^{\rm c}$ is defined by
\begin{align}
	\Gamma^{\rm c} = (K^{-1} + A)^{-1}.
\end{align}
Substituting eq.~(\ref{eq:K_Gamma_c}) into eq.~(\ref{eq:chi_c_q_K_q}), we finally obtain the equation for $\chi^{\rm c}_{\mib{q}}$ as follows:
\begin{align}
	\chi^{\rm c}_{\mib{q}} &= \Pi_{\mib{q}} + \Pi_{\mib{q}} \Gamma^{\rm c} \Pi_{\mib{q}} + \cdots \nonumber \\
	&= \Pi_{\mib{q}} + \Pi_{\mib{q}} \Gamma^{\rm c} \chi^{\rm c}_{\mib{q}}. 
\end{align}
Therefore, $\Gamma^{\rm c}$ signifies the irreducible vertex part for the conduction electrons, and is composed only of local quantities in infinite dimensions.

\end{document}